\documentclass[a4paper,fleqn,usenatbib]{mnras}
\usepackage{mathptmx}
\usepackage{txfonts}

\usepackage[T1]{fontenc}
\usepackage{ae,aecompl}

\usepackage{graphicx}	% Including figure files
\usepackage{amssymb}	% Extra maths symbols
\usepackage{subfig}
\usepackage{xcolor}
\usepackage{pdflscape}
\usepackage[framemethod=tikz]{mdframed}

\title[Resilience of Sloshing Cold Fronts]{Resilience of Sloshing Cold Fronts against Subsequent Minor Mergers}

\author[I. M. Vaezzadeh et al.]{
Iraj Vaezzadeh,$^{1}$\thanks{E-mail: I.M.Vaezzadeh-2018@hull.ac.uk},
Elke Roediger,$^{1}$
Claire Cashmore,$^{1}$
Matthew Hunt$^{1}$,
John ZuHone$^{2}$,
\newauthor
William Forman$^{2}$,
Christine Jones$^{2}$,
Ralph Kraft$^{2}$,
Paul Nulsen$^{2,3}$,
Yuanyuan Su$^{4}$,
\newauthor
Eugene Churazov$^{5,6}$
\\
$^{1}$E.A. Milne Centre, University of Hull, Cottingham Road, HU6 7RX, U.K.\\
$^{2}$Smithsonian Astrophysical Observatory, Harvard-Smithsonian Center for Astrophysics, 60 Garden St., Cambridge, MA 02138, USA\\
$^{4}$Physics and Astronomy, University of Kentucky, 505 Rose street, Lexington, KY 40506, USA\\
$^{3}$ICRAR, University of Western Australia, 35 Stirling Hwy, Crawley, WA 6009, Australia\\
$^{5}$Max Planck Institute for Astrophysics, Karl-Schwarzschild-Str 1, D-85741 Garching, Germany\\
$^{6}$Space Research Institute (IKI), Profsoyuznaya 84/32, Moscow 117997, Russia\\
}

\date{Accepted XXX. Received YYY; in original form ZZZ}

\pubyear{2020}

\begin{document}
\label{firstpage}
\pagerange{\pageref{firstpage}--\pageref{lastpage}}
\maketitle

\begin{abstract}
Minor mergers are common in galaxy clusters.
They have the potential to create sloshing cold fronts (SCFs) in the intracluster medium (ICM) of the cluster.
However, the resilience of SCFs to subsequent minor mergers is unknown.
Here we investigate the extent to which SCFs established by an off-axis minor merger are disrupted by a subsequent minor merger.
We perform a suite of 13 hydrodynamic + N-body simulations of idealised triple cluster mergers in which we vary the approach direction and impact parameter of the tertiary cluster.
Except for $\sim$1\,Gyr after the first core passage of the tertiary cluster, clear SCFs are present in all merger configurations.
Subsequent head-on minor mergers reduce the number of SCFs significantly, while subsequent off-axis minor mergers only moderately reduce the number of SCFs.
In particular, outer ($\gtrsim$500$\,$kpc) SCFs are resilient. 
The results of this work indicate that SCFs are easily formed in the course of a minor merger and are long-lived even if a further minor merger takes place. 
SCFs thus should be ubiquitous, but deriving the merger history of a given cluster based on its observed SCFs might be more complex than previously thought.
\end{abstract}


%%%%%%%%%%%%%%%%%%%%%%%%%%%%%%%%%%%%%%%%%%%%%%%%%%%%%%%%%%%%%%%%%%%%%%%%%%%%%%%%%%%%%%%%%%%%%%%%%%%%
%%%%%%%%%%%%%%%%%%%%%%%%%%%%%%%%%%%%%%%%%%%%%%%%%%%%%%%%%%%%%%%%%%%%%%%%%%%%%%%%%%%%%%%%%%%%%%%%%%%%
%%%%%%%%%%%%%%%%% BODY OF PAPER %%%%%%%%%%%%%%%%%%

\section{Introduction}
Galaxy clusters are the largest structures, at present, in the Universe which are gravitationally collapsed.
They are dominated by a dark matter halo which is filled with the X-ray emitting intracluster medium (ICM) and galaxies.
They represent, for a host of reasons, an important probe of cosmological parameters and astrophysical processes on large scales.
According to the current model of hierarchical structure formation, galaxy clusters grow via mergers with other galaxy clusters and groups, and accretion of gas and dark matter from the cosmic web \citep{Kravtsov&Borgani2012}.
Mergers of galaxy clusters are the most energetic events in the Universe since the Big Bang \citep{Sarazin2002}, driving shocks and creating turbulence in the ICM.

Mergers can occur between similarly sized ($\sim$1:1) clusters (major mergers) or between a main cluster and a smaller cluster or galaxy group (minor merger, e.g. $\sim$1:3 or less).
Minor mergers have been shown to induce contact discontinuities (cold fronts) in the ICM of the primary cluster \citep{TittleyHenriksen2005, AscasibarMarkevitch2006,ZuHone2016a}.
In contrast to a shock, a cold front is colder on the denser side of the discontinuity and the ICM pressure is continuous across the front.
Both cold fronts and shocks appear in X-ray surface brightness maps as sharp edges and are expected to be common in massive clusters because minor mergers are common \citep{Hallman2010}. 
The first example of a cold front was discovered in Abell 2142 \citep{Markevitch2000, Vikhlinin2001}, with many more being discovered since (see e.g. \citealt{GhizzardiRossettiMolendi2010}).
\citet{GhizzardiRossettiMolendi2010} estimate, under the assumption of cold fronts in cool core (CC) clusters being initiated by minor mergers, that a cluster undergoes one minor merger (at least 1:10) per 3\,Gyr.
Most cold fronts come in two classes depending on their originating mechanism: sloshing cold fronts (oscillation class), merger cold fronts (remnant class) (\cite{TittleyHenriksen2005}, see \citet{MarkevitchVikhlinin2007} for a review).
However, other mechanisms have been proposed: shock-induced cold fronts (SICFs) \citep{Birnboim2010} formed by the collision of two shocks and CFs caused by colliding inflows of gas \citep{Zinger2018}.
The remnant/merger class cold fronts (hereafter MCFs) form at the boundary between the gaseous atmospheres of the main cluster and the ram-pressure stripped gas of the subcluster \citep{Markevitch2000, Vikhlinin2001, TittleyHenriksen2005, MarkevitchVikhlinin2007}.
Sloshing cold fronts (hereafter SCFs) can arise when an off-axis minor merger perturbs the cool-core of the primary cluster (pulling it towards the passing perturber), imparting angular momentum and causing the primary cluster’s ICM to ‘slosh’ sub-sonically about the gravitational potential, producing arc-like edges wrapped around the cluster core, initially at small cluster-centric radii ($\lesssim100$\,kpc) \citep{TittleyHenriksen2005, AscasibarMarkevitch2006, OwersNulsenCouch2011} which continue to grow to large radii \citep{Simionescu2012,Rossetti2013,Walker2014}.

The presence of cold fronts in galaxy clusters has been found to be almost ubiquitous.
\citet{Markevitch2003} found that 2/3 of their sample of 37 cool-core clusters (then termed cooling flow clusters) contained cold fronts.
\citet{GhizzardiRossettiMolendi2010} subsequently found, from an X-ray flux-limited sample of 32 clusters that 19 clusters contain at least one cold front (12 of which are SCFs: Centaurus, A262, Perseus, A2199, 2A0335, A3558, A496, A1644, A2065, A3562, A1795, and A576) and that they are readily found in systems undergoing a merger in the plane of the sky.

The morphology and sharpness of cold fronts have been appreciated as a unique tool to probe the microphysics of the ICM (e.g. viscosity, turbulence, magnetic fields, etc., see \citealt{ZuHone2016a} for a review).
The interpretation of such cold fronts can be complicated by the unknown merger history of the cluster.
Binary merger simulations tailored to specific clusters have been used to determine their cluster's merger histories \citep{Roediger2011,Roediger2012b,Sheardown2018} as well as using these merger histories and ICM flow patterns to study additional cluster physics.
However, if SCFs are not only long-lived, but also resilient to subsequent minor mergers, this raises the question of how accurately a given cluster's merger history and current ICM motions can be derived from the observed sloshing pattern. 
The study of the evolution of SCFs and how resilient they are against subsequent mergers can help us to break the degeneracy.

Thus far, work in the literature has focused on idealised binary mergers to explore the nature of gas sloshing in galaxy clusters, with no work that we are aware of being focused on the effect of multiple mergers on SCFs.
Simulated binary mergers show that, left alone, the SCFs persist for extended times (>10\,Gyr) \citep{AscasibarMarkevitch2006, Roediger2012a, ZuHone2011}.
In this paper we show that the presence of SCFs is resilient to subsequent minor mergers, i.e. SCFs exist throughout the evolution except for $\sim$1\,Gyr after the tertiary cluster's infall (all times stated in this paper are relative to the simulation start time).
To this end, we run a suite of twelve hydrodynamic + N-body triple merger simulations (in addition to a binary merger for reference) exploring the effect that different trajectories of a second minor merger have on the established SCFs. 
%%%%%%%%%%%%%%%%%%%%%%%%%%%%%%%%%%%%%%%%%%%%%%%%%%%%%%%%%%%%%%%%%%%%%%%%%%%%%%%%%%%%%%%%%%%%%%%%%%%%
%%%%%%%%%%%%%%%%%%%%%%%%%%%%%%%%%%%%%%%%%%%%%%%%%%%%%%%%%%%%%%%%%%%%%%%%%%%%%%%%%%%%%%%%%%%%%%%%%%%%
%%%%%%%%%%%%%%%%%%%%%%%%%%%%%%%%%%%%%%%%%%%%%%%%%%%%%%%%%%%%%%%%%%%%%%%%%%%%%%%%%%%%%%%%%%%%%%%%%%%%

\section{Method}
\subsection{Simulations}
We run the simulations using the hydrodynamic + N-body code, FLASH v4.3 \citep{Fryxell2000}.
FLASH is a modular multi-physics Eulerian adaptive mesh refinement (AMR) hydrodynamic code.
The use of an N-body solver with high particle resolution allows us to accurately model effects like dynamical friction which captures the evolution of collisionless components (DM, galaxies), and the decay of the subcluster orbits due to dynamical friction and tidal stripping.

Our simulations run in a three-dimensional domain of 13\,Mpc$^{3}$ in size, with periodic (`wrap-around') boundary conditions, chosen due to the required initial separation of the clusters and to ensure that boundary effects do not affect the mergers.
Our simulations run for $\sim$10\,Gyr with snapshots produced every 50\,Myr.
For simplicity, we do not take account of cosmological expansion and assume a flat $\Lambda$CDM cosmology with $h=0.7$.
For the analysis we make use of the Python based library, yt \citep{Turk2011}.  

The AMR allows us to place refinement in areas of interest and save computational effort in areas of less interest.
As such we refine on particle density: when the number of particles in a block of 16$^{3}$ grid cells exceeds 1100, the block is refined; similarly if the number of particles drops below 550 the block will be derefined.
We achieve a maximum spatial resolution of 3.7\,kpc within, typically, the inner 250\,kpc for the primary cluster and the inner 125\,kpc for the secondary and tertiary clusters.
Within the primary region of interest (innermost 4\,Mpc of the simulation domain), the lowest level of resolution is 25.4\,kpc; the lowest resolution in the domain is 101\,kpc.
The simulations use $5\times10^{6}$ particles for the primary and $5\times10^{5}$ for the secondary and tertiary clusters.

%%%%%%%%%%%%%%%%%%%%%%%%%%%%%%%%%%%%%%%%%%%%%%%%%%
\subsubsection{Initial Conditions}
\begin{figure}
	\includegraphics[width=\columnwidth,trim=0cm 0cm 0cm 0cm,clip]{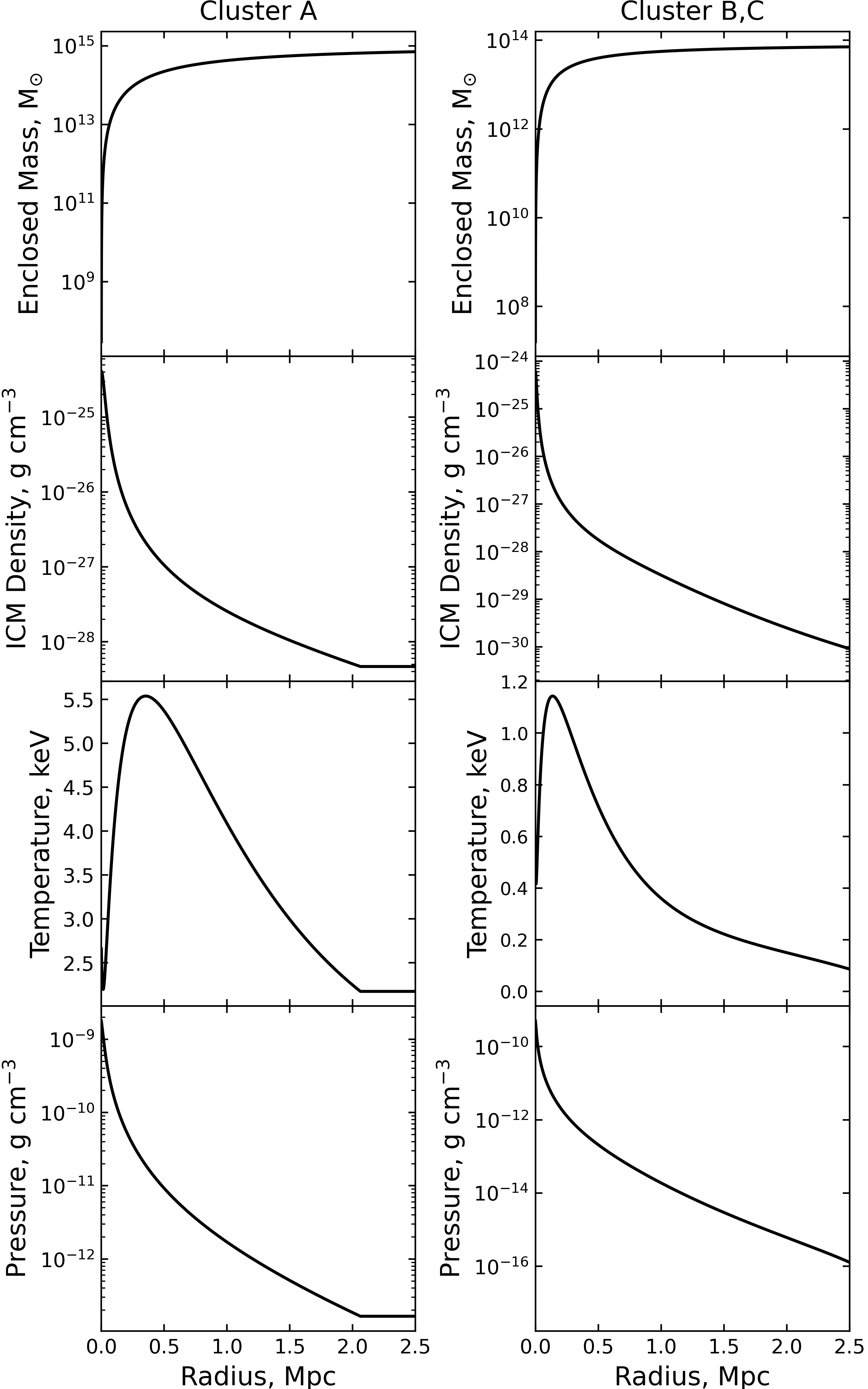}
    \caption{Radial profiles of the enclosed mass, gas density, gas temperature and gas pressure of cluster A (left) and clusters B and C (right). The density, temperature and pressure at which the WHIM-like atmosphere are set can be seen in the profiles of cluster A. }
    \label{fig:profiles}
\end{figure}

As we are not attempting to recreate a particular system, the clusters need not be tuned to any specific X-ray observation or temperature profile and can instead be generic.
Radial profiles of key cluster parameters can be seen in Figure \ref{fig:profiles}.
All clusters used in the simulations are designed to be in hydrostatic equilibrium with spherical symmetry and self-gravity, using the method of \citet{ZuHone2011}.
We refer to the primary, secondary and tertiary clusters as clusters A, B and C, respectively.
Where appropriate, clusters B and C are also referred to as first and second infaller, respectively.

The ICM is modelled as an ideal gas using the standard $\beta$ profile of \citet{Cavaliere1976}, cut at $r_{\mathrm{cut}}=2783$\,kpc to satisfy convergence:
\begin{equation}
\rho_{\mathrm{ICM}}(r)=\rho_{0} \left(1+\left(\frac{r}{r_{\mathrm{c}}}\right)^{2}\right)^{-\frac{3 \beta}{2}}    \left(1+\left(\frac{r}{r_{\mathrm{cut}}}\right)^{3}\right)^{-1}
\end{equation}
where $r_c$ is core radius. 
\citet{Mohr1999} find (based on a detailed study of an X-ray flux limited sample of 45 clusters) the approximate value of $\beta$ to be $\sim$0.64 for clusters and as such $\beta=0.64$ is the value we adopt. 

The total density of the clusters is set to a spherical Hernquist profile \citep{Hernquist1990} which matches the Navarro-Frenk-White profile within $0.1r_{200}$ but has the advantage of dropping off more steeply in density which converges to a finite total mass \citep{Springel&Hernquist&DiMatteo2005a}:

\begin{equation}
\rho_{\mathrm{total}}(r)=\frac{M_{\mathrm{total}}}{2 \pi} \frac{a}{r(r+a)^{3}}
\end{equation}
where $M_{\mathrm{total}}$ is the total mass and $a$ is the scale radius (a description of which can be found in \citet{Donnert2014Gordo}) with $a=566$\,kpc for cluster A and $a=230$\,kpc for clusters B and C (clusters B and C are identical in all respects).
Note that this is the total density of the cluster (DM and ICM) with the particle density of the cluster being set as the difference between this total Hernquist profile and the ICM $\beta$-profile.

As we initialise cluster C at a large distance from cluster A (to control when it reaches pericentre) we embed all clusters in a warm-hot intergalactic medium (WHIM)-like atmosphere to truncate the atmospheres of the in-falling clusters and to avoid unrealistic sizes.
We model the WHIM-like atmosphere with a uniform density, temperature and pressure with values of $\rho=4.67 \times10^{-29}$\,g\,cm$^{-3}$, $T=2.17$\,keV, $P=1.65 \times10^{-13}$\,erg\,cm$^{-3}$, these parameters equal the ICM parameters of cluster A at $r=2062$\,kpc, which can be seen in Figure \ref{fig:profiles}.
As our WHIM-like atmosphere is slightly warmer than the typical range found in the literature ($10^{5}-10^{7} \mathrm{~K}$, \citealt{Dave2001}) due to technical reasons, we do not draw conclusions concerning absolute temperatures within the simulations.
We have tested the binary merger simulation with a selection of WHIM parameters and find that our conclusions would not be affected by this choice.
The masses of the clusters are chosen to be representative of an average sized cluster for cluster A and a small cluster for clusters B and C.

\begin{table}
	\centering
	\caption{Cluster parameters}
	\label{tab:cluster_parameters}
	\begin{tabular}{ccccccr} % four columns, alignment for each
		\hline
		Cluster & $M_{200}(M_{\odot})$ & $r_{\mathrm{200}}$(kpc) & $N_{p}$ & Particle Mass$(M_{\odot})$\\
		\hline
		A & \(5 \times 10^{14}\) & 1637 & \(5 \times 10^{6}\) & \(1.30 \times 10^{8}\)\\
		B & \(5 \times 10^{13}\) & 760   & \(5 \times 10^{5}\) & \(1.21 \times 10^{8}\)\\
		C & \(5 \times 10^{13}\) & 760   & \(5 \times 10^{5}\) & \(1.21 \times 10^{8}\)\\
		\hline
	\end{tabular}
\end{table}

%%%%%%%%%%%%%%%%%%%%%%%%%%%%%%%%%%%%%%%%%%%%%%%%%%
\subsubsection{Setup parameters}
The two clusters that constitute the initial binary merger (A and B) are set with a distance equalling the sum of their respective $\,r_{\mathrm{200}}$ radii.
We motivate this separation based on the findings of \citet{Vitvitska2002} that, in cosmological simulations, the velocity of an in-falling subcluster is $\sim1.1\,V_{\mathrm{c}}$ at $r_{\mathrm{vir}}$ (for which we use $\,r_{\mathrm{200}}$ as a proxy), where $V_{c}^{2}=G M_{\mathrm{vir}} / R_{\mathrm{vir}}$ is the circular velocity of the Hernquist profile; for cluster A, $V_{\mathrm{c}}=1150$\,kms$^{-1}$.
\citet{Vitvitska2002} also revealed that the tangential velocity dispersion of a merger is dependent on the mass ratio of the primary cluster to the secondary.
They find that the average tangential velocity for minor mergers, is $v_{\perp}=0.71\,V_{\mathrm{c}}$, corresponding to 817\,kms$^{-1}$ in our simulations.
We control the pericentre distance (the closest approach of the two cores) of the simulations through the use of a tangential velocity component in accordance with these findings.
We find that the parameters found in Table \ref{tab:binary_parameters} (between 0\,kms$^{-1}$ (head-on) and 817\,kms$^{-1}$) produce clear sloshing fronts.

A mass ratio of R$=$1:10 is shown to create clear SCFs in the host cluster \citep{ZuHone2011} and is thus chosen as the mass ratio between the merging clusters.
For the second infaller (cluster C) we use ad hoc tangential velocities to produce, for each approach approach, an on-axis merger and one off-axis merger to each side of cluster A.
We adopt a naming convention in with head-on mergers are named HeadX, clockwise mergers ClockX and anti-clockwise mergers AntiClockX.
The parameters used for cluster C in each simulation can be found in Table \ref{tab:tertiary_parameters}.
The off-axis infalls of cluster C typically have a pericentre distance of $\sim$100\,kpc.

A further simulation probing the effect of a subsequent merger out of the plane of the initial merger has also been performed.
In this simulation (which we call LOS1) cluster C is initialised with the parameters found in Table \ref{tab:supplementary_parameters}.
Cluster C has an off-axis approach despite only having a radial velocity component due to the drift of cluster A after cluster B's infall. 

\begin{table}
	\centering
	\caption{Initial binary merger parameters. Both clusters have a $y$-position of 0, $\mathbf{v}_{\|}$ represents velocity in the $x$-direction and $\mathbf{v}_{\perp}$ represents velocity in the $y$-direction.}
	\label{tab:binary_parameters}
	\begin{tabular}{lcccr} % four columns, alignment for each
		\hline
		Cluster & x-position (kpc)  & $\mathbf{v}_{\|}$ (km/s) & $\mathbf{v}_{\perp}$ (km/s)\\
		\hline
		1 & 0  & 0 & 0 \\
		2 & 2397  & -966 & -299 \\
		\hline
	\end{tabular}
\end{table}

\begin{table}
	\centering
	\caption{Positions and impact parameters for the second infaller in each simulation. The positions and velocities of the primary cluster and first infaller are constant in every case.}
	\label{tab:tertiary_parameters}
	\begin{tabular}{lccccr} % four columns, alignment for each
		\hline
		Simulation & $x$ (Mpc) & $y$ (Mpc)  & $\mathbf{v}_{\|}$ (km/s) & $\mathbf{v}_{\perp}$ (km/s)\\
		\hline
		Head1  	  & 0		& 5		& -100	& -32.7 \\
		Head2 	  & 5		& 0		& -100	& 0 \\
		Head3 	  & 0		& -5		& 100	& -75 \\
		Head4	  & -5	& 0 		& 100 	& -100 \\
		Clock1	  & 0		& 5		& -100	& 0 \\
		Clock2	  & 5		& 0		& -100	& -175 \\
		Clock3	  & 0		& -5		& 100	& -225 \\
		Clock4        & -5 	& 0		& 100	& 0 \\	
		AntiClock1  & 0	 	& 5  		& -100	& -200 \\
		AntiClock2  & 5   	& 0 		& -100	& 32.7 \\
		AntiClock3  & 0	 	& -5		& 100	& 32.7 \\
		AntiClock4  & -5 	& 0 		& 100	& -200 \\
		\hline
	\end{tabular}
\end{table}

\begin{table}
	\centering
	\caption{Position and impact parameters for the second infaller in the supplementary simulation LOS1. The $x$ and $y$ positions are both zero. The positions and velocities of the primary cluster and first infaller are the same as in Table \ref{tab:binary_parameters}.}
	\label{tab:supplementary_parameters}
	\begin{tabular}{lccccr} % four columns, alignment for each
		\hline
		Simulation  & $z$ (Mpc) & $\mathbf{v}_{\|}$ (km/s) & $\mathbf{v}_{\perp}$ (km/s)\\
		\hline
		LOS1  & 4500 & -100 & 0 \\
		\hline
	\end{tabular}
\end{table}
%%%%%%%%%%%%%%%%%%%%%%%%%%%%%%%%%%%%%%%%%%%%%%%%%%%%%%%%%%%%%%%%%%%%%%%%%%%%%%%%%%%%%%%%%%%%%%%%%%%
\subsection{Cold Front Identification}
\label{section:CF_detection}
To quantify the number of cold fronts present in cluster A along a given direction we create a ray between two specified points in the simulation domain.
We perform the process described below separately along both the $x$ and $y$ directions.

\begin{figure}
\centering
\includegraphics[trim={0cm 0.9cm 0cm 0.5cm},clip, width=\columnwidth]{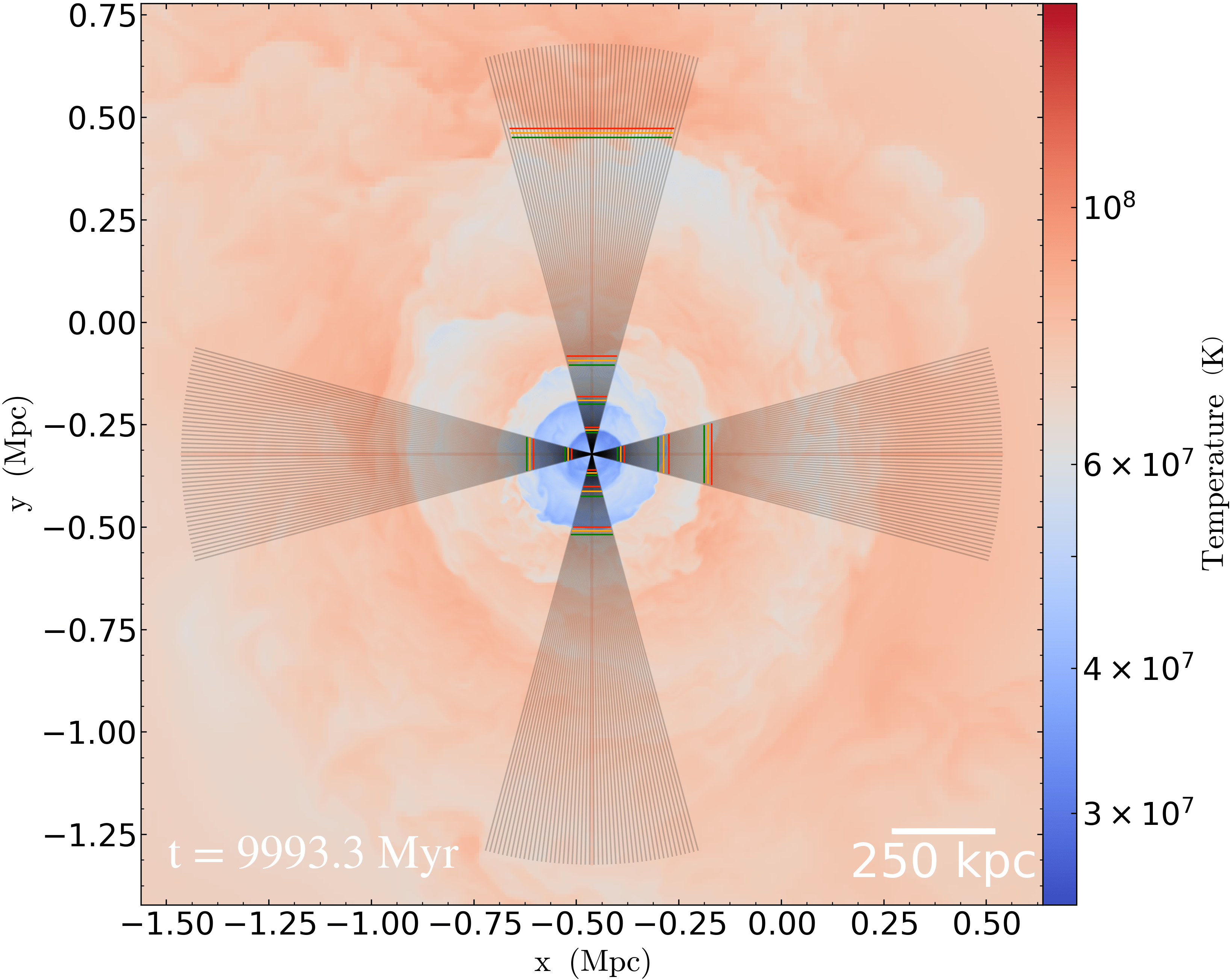} 
\caption{Temperature slice of the binary merger simulation at 10\,Gyr centred on cluster A. To calculate the averaged radial temperature profile (see Figure \ref{fig:genomsnitt_ray}), we extract and then average the profiles along the black lines that sweep 30$^{\circ}$ sectors along the $x$- and $y$-axes, with the resulting averaged ray being represented in red on this plot. The vertical green, orange and red lines denote, respectively, the start, average and end positions of identified CFs.}
\label{fig:CFdetector_sliceplot}
\end{figure}

\citet{GhizzardiRossettiMolendi2010} found that the surface brightness discontinuities in their sample span an angular extent of at least 30$^{\circ}$.
We cast 50 rays using the yt ray object (each consisting of 750 points, 1.3\,kpc apart) to form 30$^{\circ}$ sectors of a circle of radius 1\,Mpc about the centre of cluster A (defined by the minimum of its gravitational potential) which can be seen in Figure \ref{fig:CFdetector_sliceplot}.
We then take the average of all these rays to produce a single temperature within the 30$^{\circ}$ sector.

\begin{figure*}
\centering
\includegraphics[width=\textwidth]{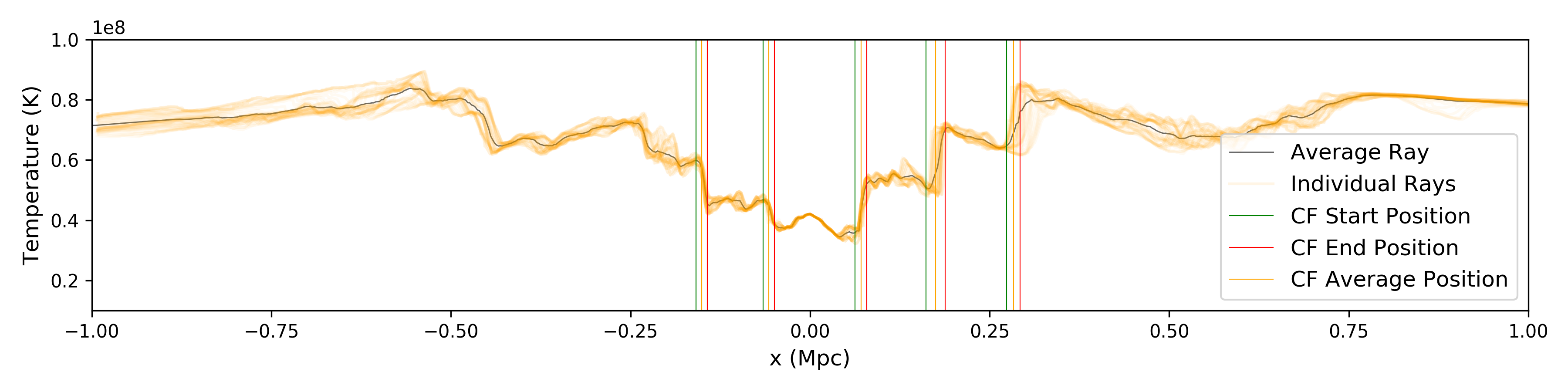} 
\caption{The averaged temperature profile in the $x$-direction of the binary merger at 10\,Gyr. Profiles along individual rays are plotted in pale orange, with their average in black. The vertical green, orange and red lines denote, respectively, the start, average and end positions of identified CFs that have met the temperature ratio threshold (>1.15). Note that the $x$-axis has been adjusted so the gravitational potential minimum (i.e. the cluster centre) is at 0\,Mpc.}
\label{fig:genomsnitt_ray}
\end{figure*}

We scan along the averaged temperature profile (seen in Figure \ref{fig:genomsnitt_ray}) and mark changes in temperature of more than 2$\%$ between neighbouring points (corresponding to a scale of 1.3\,kpc).
We repeat this process in the opposite direction to mark steep gradients on the opposite side of the core).
In some cases not all points along a steep drop will be marked as CF points in the initial sweeps due to small scale fluctuations.
To counter this, and ensure that each CF is continuous, we perform a rolling average of every three points along the profile (corresponding to a scale of $\sim$4\,kpc).
This smoothing step has the consequence of effectively setting the minimum separation between CFs as $\sim$4\,kpc.
We define the start point of each CF as the first point in the profile that has met the 2$\%$ change criterion and the end point as the last in a set of points after the smoothing.
The average CF position is simply the average of the start and end position of the CF. 

To avoid spurious detections we remove detections with a temperature ratio between the start and end points of <1.15 (based on the temperature ratio of 1.15 found in Abell 2219 by \citealt{Canning2017}).
Examples of identified SCFs can be seen in Figure \ref{fig:CFdetector_sliceplot}.
The algorithm makes no distinction as to the origin and nature of the front counted, and therefore will detect MCFs when the infalling subcluster travels through a detection sector, we therefore refer to detections while the infalling subcluster's atmosphere could be causing the identification simply as CFs rather than SCFs.
Several parameters affect the number of identified cold fronts, such as limits on gradient, minimum allowed temperature ratio or rolling average length/minimum separation.
We tested a range of these parameters on the binary merger run and visually confirmed the CF detections in temperature slice images such as Figure \ref{fig:CFdetector_sliceplot} as well projected temperature weighted by density squared such as in Figure \ref{fig:TempDens2_Movie_2,5Mpc}.
Relaxing the gradient limit and temperature ratio limit results in more CFs being identified, but such weaker CFs would not be easily identifiable in observations.
Our conclusions regarding the resilience of SCFs do not depend on the exact choice of these parameters.

%%%%%%%%%%%%%%%%%%%%%%%%%%%%%%%%%%%%%%%%%%%%%%%%%%%%%%%%%%%%%%%%%%%%%%%%%%%%%%%%%%%%%%%%%%%%%%%%%%%%
%%%%%%%%%%%%%%%%%%%%%%%%%%%%%%%%%%%%%%%%%%%%% RESULTS %%%%%%%%%%%%%%%%%%%%%%%%%%%%%%%%%%%%%%%%%%%%%%%%%
%%%%%%%%%%%%%%%%%%%%%%%%%%%%%%%%%%%%%%%%%%%%%%%%%%%%%%%%%%%%%%%%%%%%%%%%%%%%%%%%%%%%%%%%%%%%%%%%%%%%
\section{Results} 
\subsection{Sloshing process}

We briefly describe the mechanism which causes the ICM sloshing in the simulations, making reference to Figure \ref{fig:Binary_movie}, which shows the binary merger simulation.
Note that the colour scale has been made constant for all simulations so that comparisons may be more readily drawn.
This has the consequence that saturation occurs at certain times/locations such as when gas is shock heated, but the features we are interested in are unaffected.
In all simulations, the sloshing patterns caused by the initial binary merger between cluster A and cluster B are formed in the same way.

Cluster B approaches the core of cluster A, pulling the dark matter and ICM peaks of cluster A equally towards cluster B.
The ICM peak of cluster A is subject to ram pressure, unlike the DM peak, and as such is no longer in equilibrium with the DM-peak.
When cluster B passes its pericentre position ($\sim$200\,kpc) at $\sim$1.5\,Gyr, the gas core of cluster A is no longer subject to this ram pressure, and begins to fall from this maximum separation back to the gravitational potential minimum, coinciding with the DM peak.
As the gas core falls back, it ’overshoots’ the DM peak and travels to a new maximum position on the opposite side.
In this fashion, the gas continues to oscillate about the DM peak of cluster A, creating the characteristic ’sloshing’ fronts staggered about the core of cluster A. 
Cluster B passes south of cluster A, attracting its DM core to the south, thereby initiating the sloshing in a north-south direction.
Cluster B’s off-axis infall trajectory is such that it imparts angular momentum to cluster A and causes cluster A to ’swirl’ in a clockwise direction after first pericentric passage ($\sim$1.5\,Gyr) which gives the resultant spiralling SCFs a clockwise direction.

Cluster B reaches first apocentre at $\sim$2.75\,Gyr; at this stage the ICM in the core of cluster A is clearly sloshing about the perturbed potential and creating SCFs.
Second pericentric passage occurs at $\sim$4.1\,Gyr; other than the CFs that are directly hit by the infalling subcluster, the overall pattern of SCFs remains resilient to this second pericentric passage.
The sloshing fronts then continue to travel outwards, for at least 8.5\,Gyr (until $t_{\mathrm{max}}$), from the core, with subsequent pericentric passages of cluster B not affecting their evolution.
Several SCFs are visible at $t_{\mathrm{max}}$, forming a clear (though not perfectly continuous) spiral shape out to $\sim$800\,kpc, which is $\sim$0.5$\,r_{\mathrm{200}}$.
SCFs are clearly present at $\sim$5\,Gyr in Figure \ref{fig:Binary_movie}, which is the approximate time that cluster C will reach first pericentre.

%%%%%%%%%%%%%%%%%%%%%%%%%%%%%%%%%%%%%%%%%%% TRIPLE MERGERS %%%%%%%%%%%%%%%%%%%%%%%%%%%%%%%%%%%%%%%%%%%%%%
%%%%%%%%%%%%%%%%%%%%%%%%%%%%%%%%%%%%%%% QUALITATIVE DESCRIPTION PREFACE %%%%%%%%%%%%%%%%%%%%%%%%%%%%%%%%%%%%%%%%%%%
\subsection{Evolution of Triple Mergers - Qualitative Description}
\label{section:visual_section}
\begin{figure}
	\centering
	\includegraphics[width=0.75\columnwidth]{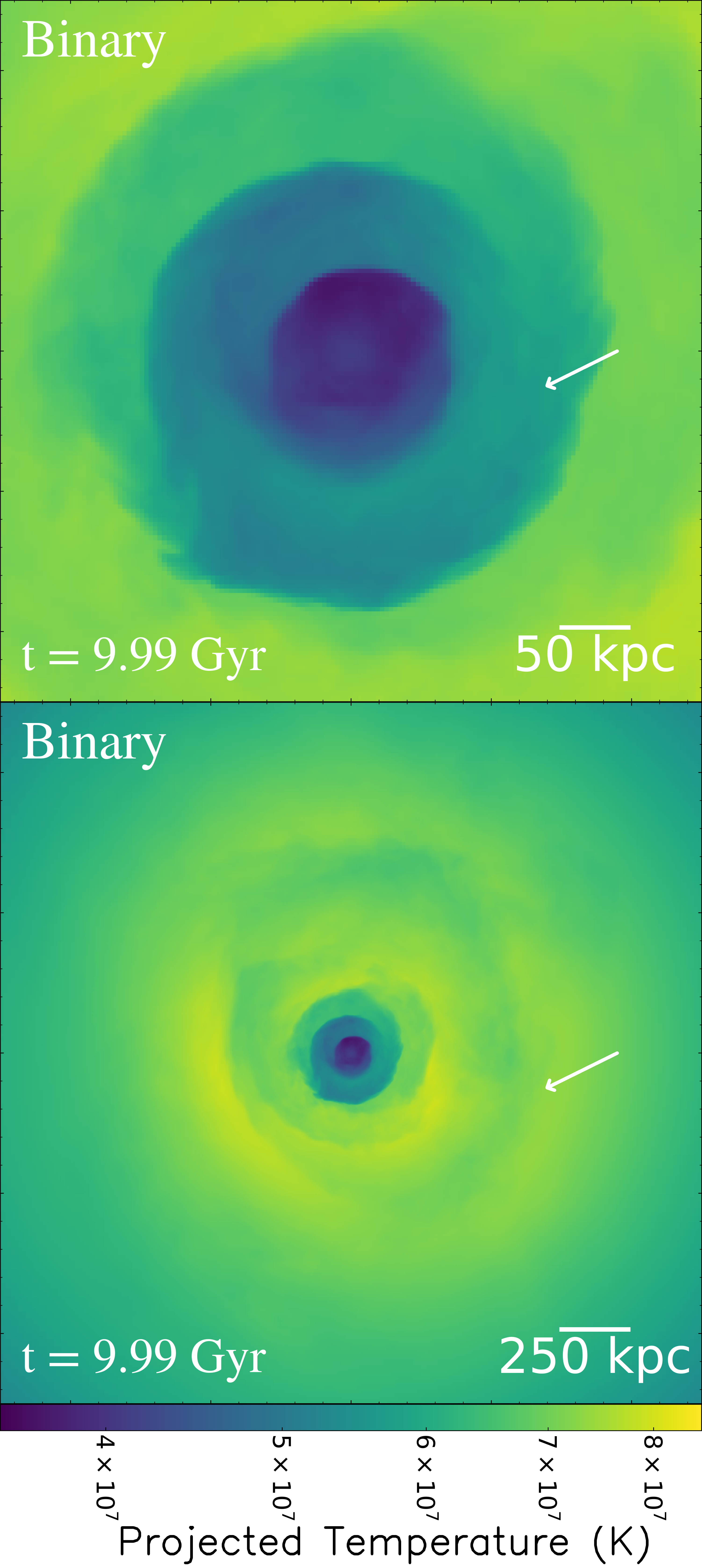} % this is a placeholder image for the movie Binary.mov
    \caption{Movie (the movie can be viewed at https://youtu.be/fZYljFAQfBA) of projections of temperature with density squared weighting through the $x-y$ plane of the binary merger simulation. The top panel is 500\,kpc on a side, and the bottom panel is 2.5\,Mpc on a side. The arrow denotes the trajectory of the infaller (note this is not to scale and serves only as a guide).}
    \label{fig:Binary_movie}
\end{figure}

To draw meaningful comparisons between simulations, we first qualitatively describe the evolution of the triple mergers based on visual inspection.
We use slices of temperature and projections of temperature weighted by density squared (to better represent observational images) to analyse the simulations.
All images are viewed with the same colour scale and field of view, which in some cases causes saturation.
It is the effect of a subsequent minor merger on the SCFs established by the initial off-axis binary minor merger that we seek to determine.
As all simulations proceed in much the same way until cluster C's first pericentric passage (see Figure \ref{fig:Binary_movie} until $\sim$5.3\,Gyr), we begin analysis of the triple mergers at this stage.
We have made the decision to keep the position of cluster C constant for each approach direction, and as such the first pericentre time (5.3-5.9\,Gyr) and distance (50-100\,kpc) of cluster C varies across the simulations due to the drift of cluster A after the infall of cluster B.

\begin{landscape}
\begin{figure}   
   \includegraphics[width=\textheight ]{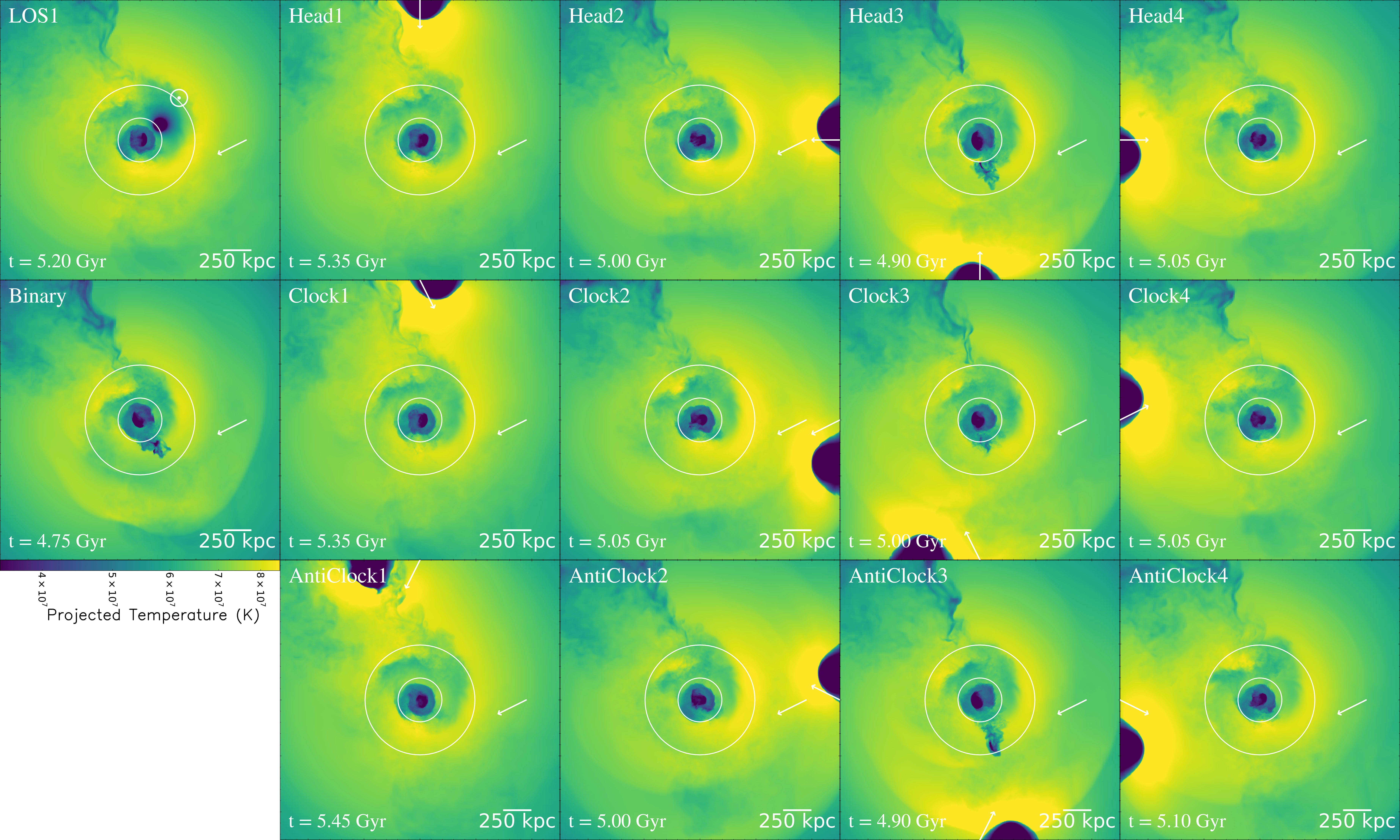} 
    \caption{Projections of temperature with density squared weighting through the $x-y$ plane for all simulations just prior to first pericentre of cluster C. The phase of each simulation has been shifted such that the 2nd infaller reaches first pericentre simultaneously in Figures \ref{fig:1st_snapshot}-\ref{fig:last_snapshot}. Each panel is 2.5\,Mpc on a side. The innermost white circle common to all panels marks the boundary of the region we define as small scale (less than 0.12$\,r_{\mathrm{200}}$ ($\sim$200\,kpc)). The outermost white circle common to all panels marks the boundary of the region we define as intermediate (between 0.12 and 0.3$\,r_{\mathrm{200}}$). Beyond this outermost circle is the large scale region (greater than 0.3$\,r_{\mathrm{200}}$ ($\sim$500\,kpc)). These annotations share the same meaning across Figures \ref{fig:1st_snapshot}-\ref{fig:last_snapshot}.  }
    \label{fig:1st_snapshot}
\end{figure}
\end{landscape}

\begin{landscape}
\begin{figure}
   \includegraphics[width=\textheight ]{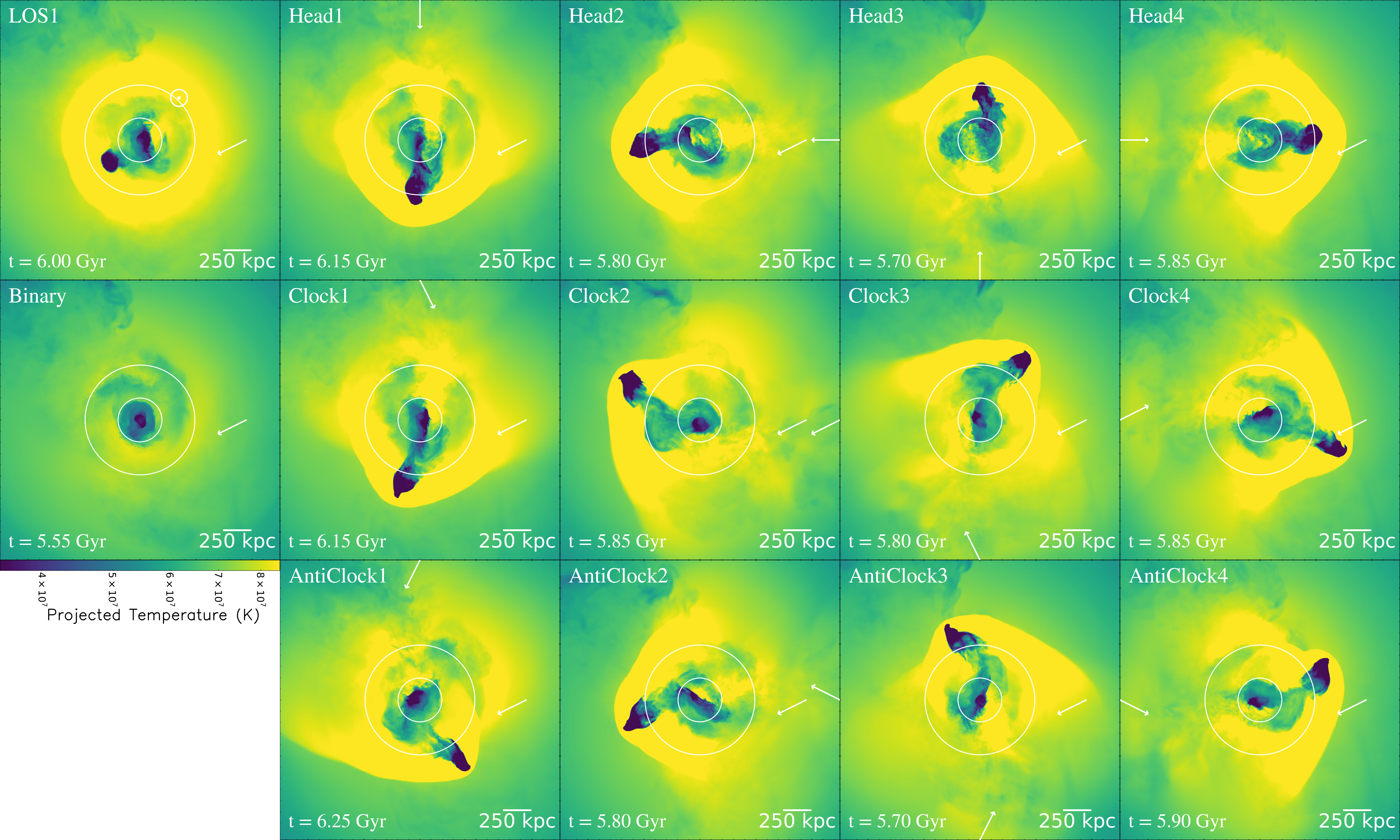} 
    \caption{Projections of temperature with density squared weighting through the $x-y$ plane for all simulations just after first pericentre of cluster C. Explanations of the annotations can be found in Figures \ref{fig:TempDens2_Movie_2,5Mpc} and \ref{fig:1st_snapshot}. In all simulations, the sloshing is highly disturbed during this phase between cluster C's first pericentre and first apocentre. Although some SCFs can be detected, they lack the characteristic spiral pattern. MCFs can be detected at the boundary of cluster C's atmosphere with the atmopshere of cluster A.}
    \label{fig:2nd_snapshot_disturbed}
\end{figure}
\end{landscape}

\begin{landscape}
\begin{figure}
   \includegraphics[width=\textheight ]{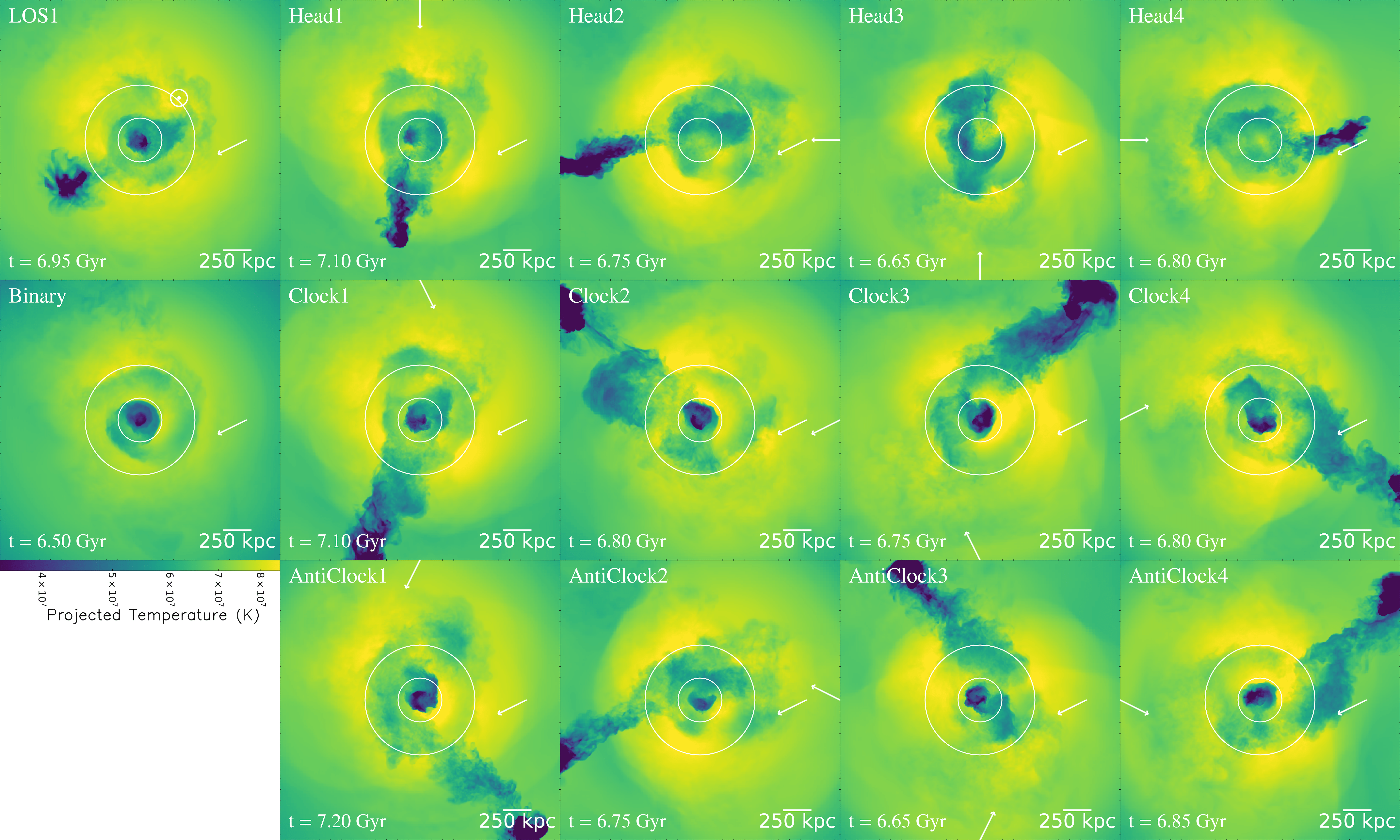} 
    \caption{Projections of temperature with density squared weighting through the $x-y$ plane for all simulations at first apocentre of cluster C. Explanations of the annotations can be found in Figures \ref{fig:TempDens2_Movie_2,5Mpc} and \ref{fig:1st_snapshot}. We see in the head-on simulations that the MCF and slingshot tail of cluster C is less pronounced than in the off-axis mergers due to the loss of much of its cool gas during its core crossing. We see new sloshing initiated in the core of cluster A in the off-axis mergers which is not present in the head-on mergers.}
    \label{fig:3rd_snapshot_apocentre}
\end{figure}
\end{landscape}

\begin{landscape}
\begin{figure}
   \includegraphics[width=\textheight ]{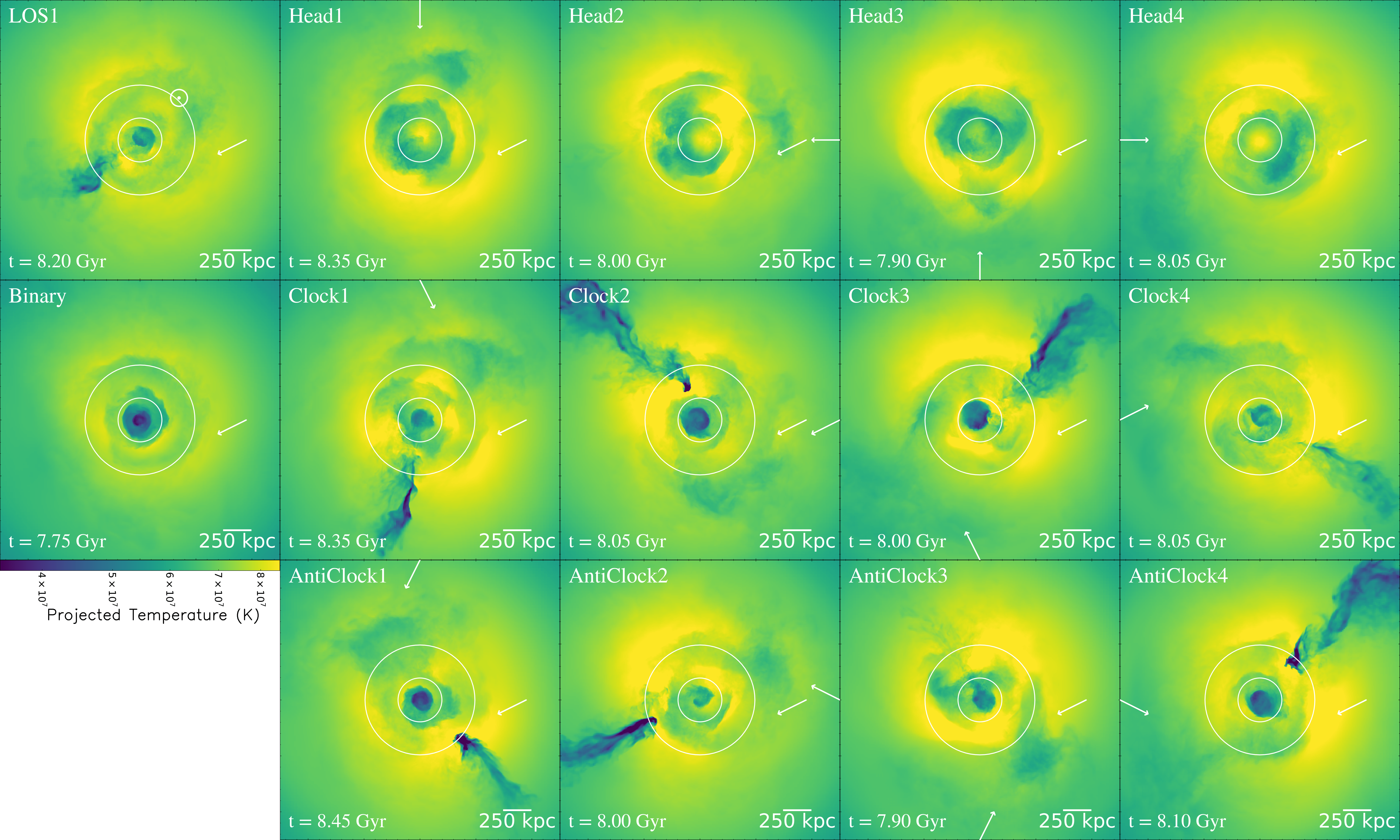} 
    \caption{Projections of temperature with density squared weighting through the $x-y$ plane for all simulations at the approximate time of second pericentre of cluster C. Explanations of the annotations can be found in Figures \ref{fig:TempDens2_Movie_2,5Mpc} and \ref{fig:1st_snapshot}. The simulations are now out of phase with regard to the second pericentre of cluster C due to the different configurations. We see that in the head-on mergers cluster C has been stripped of all of its cool gas before reaching second pericentre and is therefore not visible in this image. In the case of the off-axis mergers cluster C still carries some cool gas to second pericentre, but this has little effect on the overall evolution of the system other than to cause small-scale instabilities.}
    \label{fig:4th_snapshot_2ndperi}
\end{figure}
\end{landscape}

\begin{landscape}
\begin{figure}
   \includegraphics[width=\textheight ]{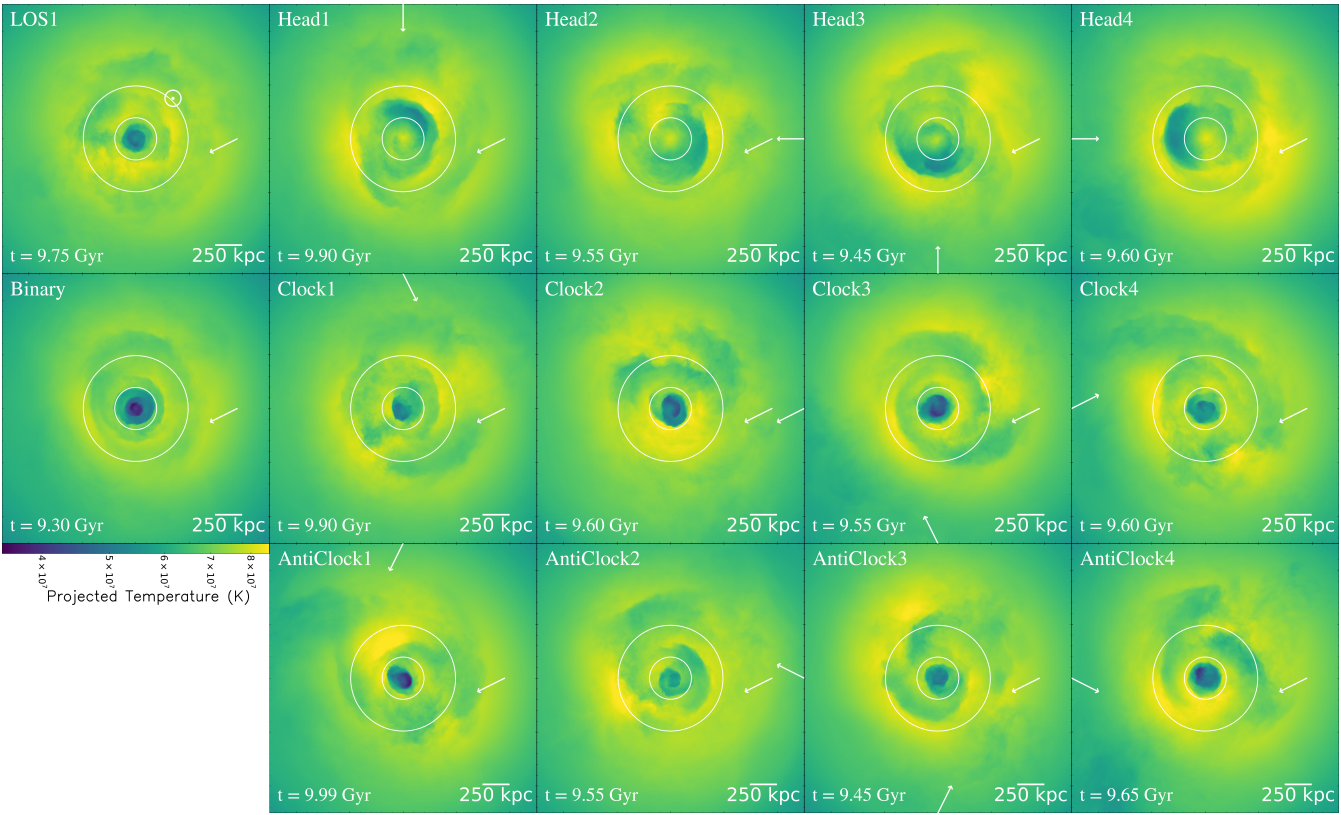} 
    \caption{Projections of temperature with density squared weighting through the $x-y$ plane for all simulations during the `relaxation phase'. Explanations of the annotations can be found in Figures \ref{fig:TempDens2_Movie_2,5Mpc} and \ref{fig:1st_snapshot}. This figure shows the state of each simulation during the `relaxation phase'; due to the difference in phase between the simulations in this figure, it represents the approximate maximum simulation time. There is a clear lack of a CC in the head-on mergers, which is likely the reason that no small scale SCFs have been generated. We see small scale SCFs in the cores of cluster A in the off-axis simulations initiated by cluster C, and large scale fronts initiated by cluster B. There is a lack of continuity between these as the intermediate scale fronts are, in some cases, destroyed, and in other cases disturbed.}
    \label{fig:last_snapshot}
\end{figure}
\end{landscape}

Due to the difference in evolution between the head-on mergers and those clockwise and anti-clockwise with respect to the initial binary merger, we describe these evolutions separately below.
In this description, we will refer to the movie (Figure \ref{fig:TempDens2_Movie_2,5Mpc}) and Figures \ref{fig:1st_snapshot}-\ref{fig:last_snapshot} in the body of the text which show the simulations aligned on the 2nd infaller's pericentre time.
The appendix features movies of each subset of simulations at different scales as well as a movie of all simulations at a scale of 500\,kpc and an image of the final time of all simulations. 

%%%%%%%%%%%%%%%%%%%%%%%%%%%%%%%%%%%%%%%%%% OVERVIEW OF TRIPLES %%%%%%%%%%%%%%%%%%%%%%%%%%%%%%%%%%%%%%%%%%%%%
\begin{landscape}
\begin{figure}
%   \newgeometry{margin=0.1mm} % and smaller margins
   \includegraphics[width=\textheight ]{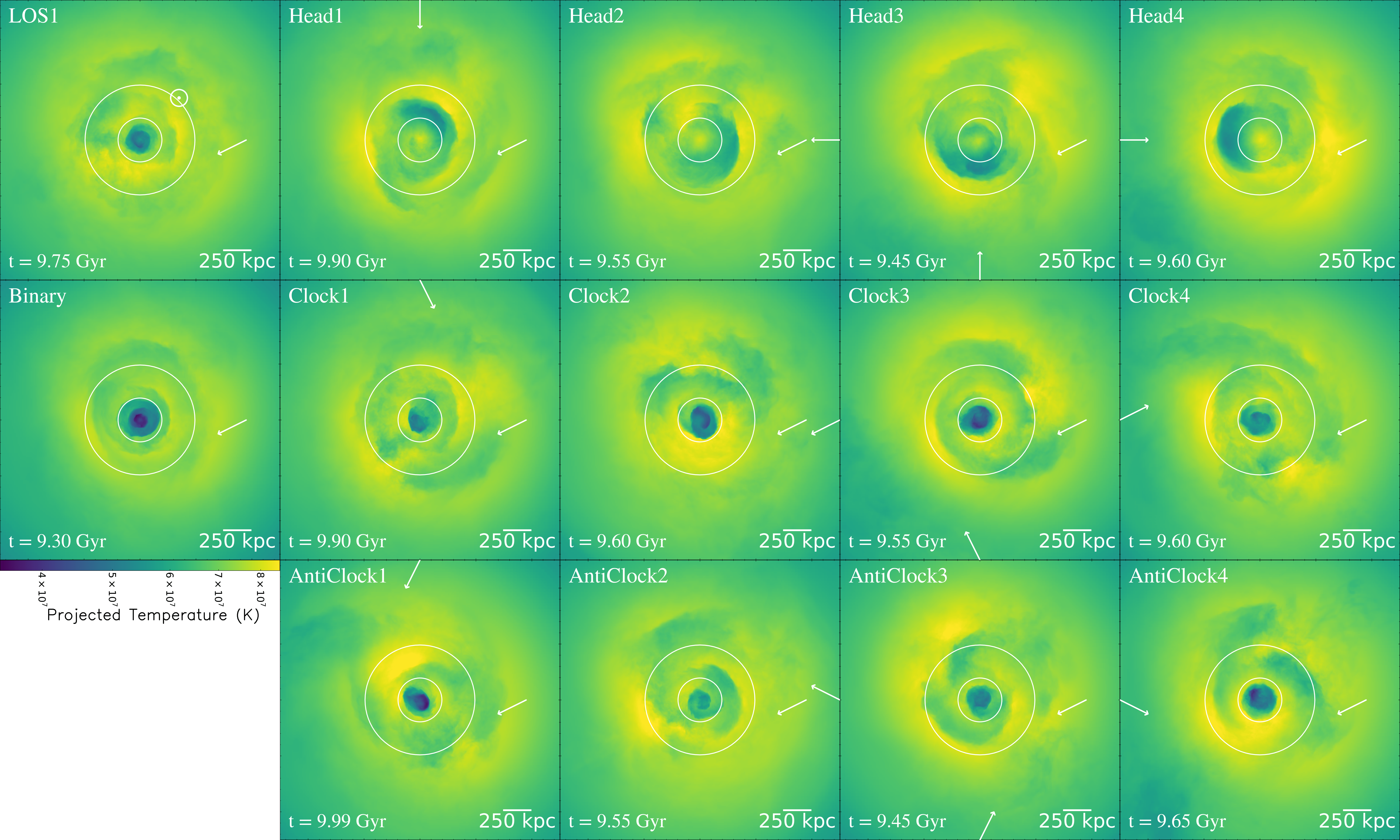} %this is a placeholder image for 2,5Mpc_2ndInfall_Regions.mov
    \caption{Movie (the movie can be viewed at https://youtu.be/97yOHejlpyI)of projections of temperature with density squared weighting through the $x-y$ plane for all simulations. Each panel is 2.5\,Mpc on a side. The rows are comprised, respectively, of the head-on, clockwise and anti-clockwise simulations; the columns represent approach direction of cluster C. The leftmost column is the binary merger and supplementary simulation (discussed in Sec \ref{section:space_sampling}) with the colour scale for all simulations. The innermost arrow in each panel represents the trajectory of the first infaller, the dotted circle in LOS1 indicates a trajectory towards the viewer of the second infaller, and the outer arrow represents the trajectory of the second infaller (note that neither arrow is to scale, and should serve only as a guide to the panels). The annotated arrows bear this meaning in all subsequent figures. It is clear from row one that a second infaller with zero impact parameter will disrupt the cool-core of the system; however, the large scale sloshing fronts remain. There is no obvious difference between the different approach directions.}
    \label{fig:TempDens2_Movie_2,5Mpc}
\end{figure}
\end{landscape}

%%%%%%%%%%%%%%%%%%%%%%%%%%%%%%%%%%%%%%%%%% SNAPSHOTS %%%%%%%%%%%%%%%%%%%%%%%%%%%%%%%%%%%%%%%%%%

 %%%%%%%%%%%%%%%%%%%%%%%%%%%%%%%%%%%%%%%%%% OVERVIEW OF TRIPLES %%%%%%%%%%%%%%%%%%%%%%%%%%%%%%%%%%%%%%%%%%%%%
\subsubsection{General overview of triple mergers}
First pericentric passage of cluster C occurs between 5.3-5.9\,Gyr.
We will refer, throughout our analysis, to specific moments (such as first pericentric passage etc.) which can be seen in Figures \ref{fig:1st_snapshot}-\ref{fig:last_snapshot}. 
SCFs extending up to $\sim$400\,kpc caused by cluster B are well established by the time of cluster C's first pericentric passage (Figure \ref{fig:1st_snapshot}).
We see strong disruption or even destruction of SCFs during the epoch from first pericentric passage until first apocentre (6.5-7\,Gyr) of cluster C (Figure \ref{fig:2nd_snapshot_disturbed}).
By the time of cluster C's first apocentre (Figure \ref{fig:3rd_snapshot_apocentre}), these SCFs initiated by the initial merger are re-established to varying degrees in the different simulations.
SCFs continue to emerge from the core of cluster A for off-axis infalls of cluster C, likely generated due to the core motion initiated by cluster C as opposed to a continuation of the sloshing caused by cluster B.
We do not distinguish between SCFs generated by cluster B and cluster C as we are interested in the number of observable SCFs after two mergers.
Cluster C's second pericentric passage (Figure \ref{fig:4th_snapshot_2ndperi}) has little effect on the evolution of the system.
We define the epoch post second pericentric passage (7.5-8\,Gyr, (Figure \ref{fig:last_snapshot})) as the `relaxation phase', which roughly corresponds to the final 2-2.5\,Gyr of the simulations.
Once the simulations enter the `relaxation phase' we typically see between 3-5 SCFs in the non-head-on mergers and 1-3 in head-on mergers.
To quantify the differences in cold front structure between simulations, we define three categories of CFs according to their position as a fraction of cluster A's $\,r_{\mathrm{200}}$.
These categories are as follows:
small scale: less than 0.12 $\,r_{\mathrm{200}}$ ($\sim$200\,kpc);
intermediate: between 0.12 and 0.3 $\,r_{\mathrm{200}}$;
large scale: greater than 0.3 $\,r_{\mathrm{200}}$ ($\sim$500\,kpc). 
These regions have been annotated on Figures \ref{fig:1st_snapshot}-\ref{fig:last_snapshot}.

 %%%%%%%%%%%%%%%%%%%%%%%%%%%%%%%%%%%%%%%%%%%%% HEAD-ONS %%%%%%%%%%%%%%%%%%%%%%%%%%%%%%%%%%%%%%%%%%%%%%%%%%
\subsubsection{Head on mergers}  \label{section:HeadOnVisual}
A movie of the evolution of the head-on mergers from cluster C's first infall, at two scales, is presented in Appendix \ref{section:Appendix A} Figure \ref{fig:HeadOn_Movie}.
When cluster C passes through the main cluster head-on, the SCFs, established by the previous merger, are thoroughly disturbed and hard to detect for the next $\sim$2\,Gyr until the `relaxation phase’.
SCFs can be picked out during this `messy' phase between the second infaller's first pericentre and second apocentre, but they lack the characteristic spiral pattern as they are so disturbed.
Detection of any SCFs between first pericentric passage ($\sim$5.5\,Gyr) and apocentre ($\sim$6.6\,Gyr) of cluster C is difficult (Figure \ref{fig:2nd_snapshot_disturbed}).
However, by the time of first apocentre (Figure \ref{fig:3rd_snapshot_apocentre}) it becomes possible to see SCFs that had already made their way out of the inner $\sim$200\,kpc.
Due to cluster C hitting the core of cluster A in a head-on fashion, cluster C loses much of its initial cool gas, hence we see a less pronounced MCF for cluster C beyond first pericentric passage.

Subsequent pericentric passages ($\sim$7.8\,Gyr and $\sim$9\,Gyr) of cluster C have a small impact on the evolution of SCFs (Figure \ref{fig:4th_snapshot_2ndperi}), making them appear more turbulent and less orderly; however, they continue to grow outwards until $t_{\mathrm{max}}$.
By approximately second apocentre ($\sim$8.3\,Gyr) the turbulence and highly disturbed morphology have decreased sufficiently so that cold fronts are detectable beyond $\sim$200\,kpc.
During the ’relaxation phase’ it is clear that the cool core of the primary cluster has been disrupted by the head-on merger.
We confirm this by viewing slices of entropy, which confirm that the entropy in the core of cluster A in these simulations is greater than in the off-axis cases. 
This approximately isentropic core is likely responsible for the lack of new SCFs emerging after pericentric passage of cluster C.
The structure of the SCFs during the ’relaxation phase’ (Figure \ref{fig:last_snapshot}), until $t_{\mathrm{max}}$, is disturbed, but a staggered spiral pattern can be seen outside the inner $\sim$200\,kpc, albeit broken in several places. 
There are fewer SCFs at $t_{\mathrm{max}}$ in all four head-on simulations than in the case of a single binary merger due to the lack of a CC resulting in no SCFs emerging from the core (Figure \ref{fig:last_snapshot}).

 %%%%%%%%%%%%%%%%%%%%%%%%%%%%%%%%%%%%%%%%%%%%% CLOCKWISE %%%%%%%%%%%%%%%%%%%%%%%%%%%%%%%%%%%%%%%%%%%%%%%%%%
\subsubsection{Clockwise mergers}  
We now turn to the simulations in which cluster C approaches cluster A in a clockwise direction with respect to our standard viewing direction, in which cluster B has a clockwise infall.
A movie of the full evolution of the clockwise mergers at two scales is presented in Appendix \ref{section:Appendix A} Figure \ref{fig:Clockwise_Movie}.
First pericentric passage ($\sim$5.6\,Gyr) disturbs the SCFs considerably and the potential is offset which causes violent gas motions in the core (Figure \ref{fig:2nd_snapshot_disturbed}).
Clock1 and Clock4 have a closer approach at first pericentric passage than Clock2 and Clock3.
This is due to the infall of cluster B on cluster A causing the system to drift, thereby misaligning the starting position of cluster C which we keep fixed for each approach direction.
This closer pericentric distance has the effect of causing more disruption to the established SCFs than is seen in Clock2 and Clock3.
SCFs can be detected during this complex phase after first pericentre, but the spiral pattern is unrecognisable in the immediate aftermath of first pericentric passage (Figure \ref{fig:2nd_snapshot_disturbed}).
SCFs emerge from the core of cluster A as cluster C reaches first apocentre ($\sim$1\,Gyr after first pericentric passage, Figure \ref{fig:3rd_snapshot_apocentre}).
The axis of sloshing is perpendicular to the infall direction of cluster C, strongly indicating that the SCFs emerging from the core beyond this time are initiated by cluster C as opposed to cluster B.
The sloshing pattern is broken in several places, and the SCFs are prone to instabilities and turbulence, with several exhibiting Kelvin-Helmholtz instabilities (KHIs).

Cluster C has maintained much of its cool gas to first apocentre ($\sim$6.8\,Gyr), and thus can be seen to exhibit a MCF and slingshot tail \citep{Sheardown2019} at first apocentre ($\sim$1.5\,Mpc from the core).
Cluster C carries very little cool gas into the core during its second pericentric passage ($\sim$8\,Gyr, Figure \ref{fig:4th_snapshot_2ndperi}); its second pericentric passage disturbs the potential causing the SCFs to shift positions, but they remain intact and continue to grow.
During the `relaxation phase' (Figure \ref{fig:last_snapshot}) there is a staggered SCF structure in the inner 300\,kpc (400\,kpc in the case of Clock1) and the large scale SCFs at $\sim$500 and $\sim$750\,kpc are still present, though they suffer varying levels of disruption and instability across the four simulations (in particular, in Clock1 the largest scale SCF is punctured by the infall of cluster C).
In all cases there are fewer SCFs at $t_{\mathrm{max}}$ than in the binary simulation, but there are still clear SCFs in the core and at large radii; clear SCFs at intermediate radii are absent.

 %%%%%%%%%%%%%%%%%%%%%%%%%%%%%%%%%%%%%%%%%%% ANTI-CLOCKWISE %%%%%%%%%%%%%%%%%%%%%%%%%%%%%%%%%%%%%%%%%%%%%%%%
\subsubsection{Anti-clockwise mergers}   
A movie of the full evolution of the anti-clockwise mergers at two scales is presented in Appendix \ref{section:Appendix A} Figure\ref{fig:AntiClock_Movie}.
For anti-clockwise infalls of cluster C (contrasting cluster B’s clockwise infall with respect to our standard viewing direction), first pericentric passage ($\sim$5.6\,Gyr) disrupts the SCFs established by the binary merger (Figure \ref{fig:2nd_snapshot_disturbed}).
There are SCFs in the core region as cluster C reaches first apocentre ($\sim$6.7\,Gyr, Figure \ref{fig:3rd_snapshot_apocentre}), seemingly initiated by the passage of cluster C as the sloshing motion is perpendicular to the infall direction of cluster C.
As cluster C reaches first apocentre (Figure \ref{fig:3rd_snapshot_apocentre}), the characteristic spiral pattern of the SCFs has appeared (though it is of course more fractured than we see in the binary merger) due to the swirling ICM combined with the sloshing gas core.
Cluster C has carried much of its cool gas to first apocentre and therefore exhibits an MCF and slingshot tail at this radius ($\sim$1.5\,Mpc from the core).

Cluster C does not carry much cool gas to second pericentre ($\sim$8\,Gyr, Figure \ref{fig:4th_snapshot_2ndperi}), and as such only its gravitational influence is felt when it reaches second pericentre.
Its second pericentric passage displaces the potential causing more sloshing along the axis perpendicular to the infall direction; however, this does not disrupt the evolution of the fronts to a large degree.
During the ’relaxation phase’ (Figure \ref{fig:last_snapshot}) after second pericentric passage there is a clear spiral structure to the SCFs within the inner 200-400\,kpc as well as large scale fronts at $\sim$700\,kpc (in the case of AntiClock1, these large scale fronts look very weak).
In all cases there are fewer SCFs during the ’relaxation phase’ and at $t_{\mathrm{max}}$ than in the binary simulation, though similarly to the clockwise set of simulations, there are still fronts at small and large radii, with intermediate fronts being those that do not survive the merger with cluster C.

%%%%%%%%%%%%%%%%%%%%%%%%%%%%%%%%%%%%%%%%%%%%%%%%%%%%%%%%%%%%%%%%%%%%%%%%%%%%%%%%%%%%%%%%%%%%%%%%%%%%
%%%%%%%%%%%%%%%%%%%%%%%%%%%%%%%%%%%%%%%% CF Counting %%%%%%%%%%%%%%%%%%%%%%%%%%%%%%%%%%%%%%%%%%%%
\subsection{Automated CF Counting}
The visual inspection presented above indicated that the infall of cluster C reduced the number of SCFs compared to the binary merger at all times and that the head-on infalls of cluster C cause the strongest reduction.
In this section, we verify this visual impression with an automated algorithm to count CFs (see \ref{section:CF_detection} for the description of the algorithm).
We count in 30$^{\circ}$ sectors along the $x$ and $y$ axes, and as such any CF with an angular extent large enough that it lies in two sectors will be counted in both tallies.
The algorithm makes no distinction as to the origin and nature of the front counted, and therefore will detect MCFs when the infalling subcluster travels through a detection sector.

Calibration of the algorithm suggests that periods during which a subcluster travels through the detection sector will see increased detection of merger class fronts.
Usually, a MCF is identified in clusters as the leading edge of the cold gaseous atmosphere of an infalling subcluster.
When a subcluster travels through the detection sector, our algorithm detects not only this edge but several behind it as well which we will consider MCFs.
This could be mitigated by running the algorithm only along the direction perpendicular to the trajectory of the infaller.
However, as we have two infallers this will quickly become impractical, and thus analysis of the triple mergers is limited to the post-second pericentric passage (`relaxation') phase of the triple merger simulations to increase confidence that the detections are purely SCFs.

 %%%%%%%%%%%%%%%%%%%%%%%%%%%%%%%%%%%%%%%%%%%%% BINARY %%%%%%%%%%%%%%%%%%%%%%%%%%%%%%%%%%%%%%%%%%%%%%%%%%
\subsubsection{Binary Merger}
\begin{figure}
	\includegraphics[width=\columnwidth,trim=0.5cm 0.1cm 1.2cm 1cm,clip]{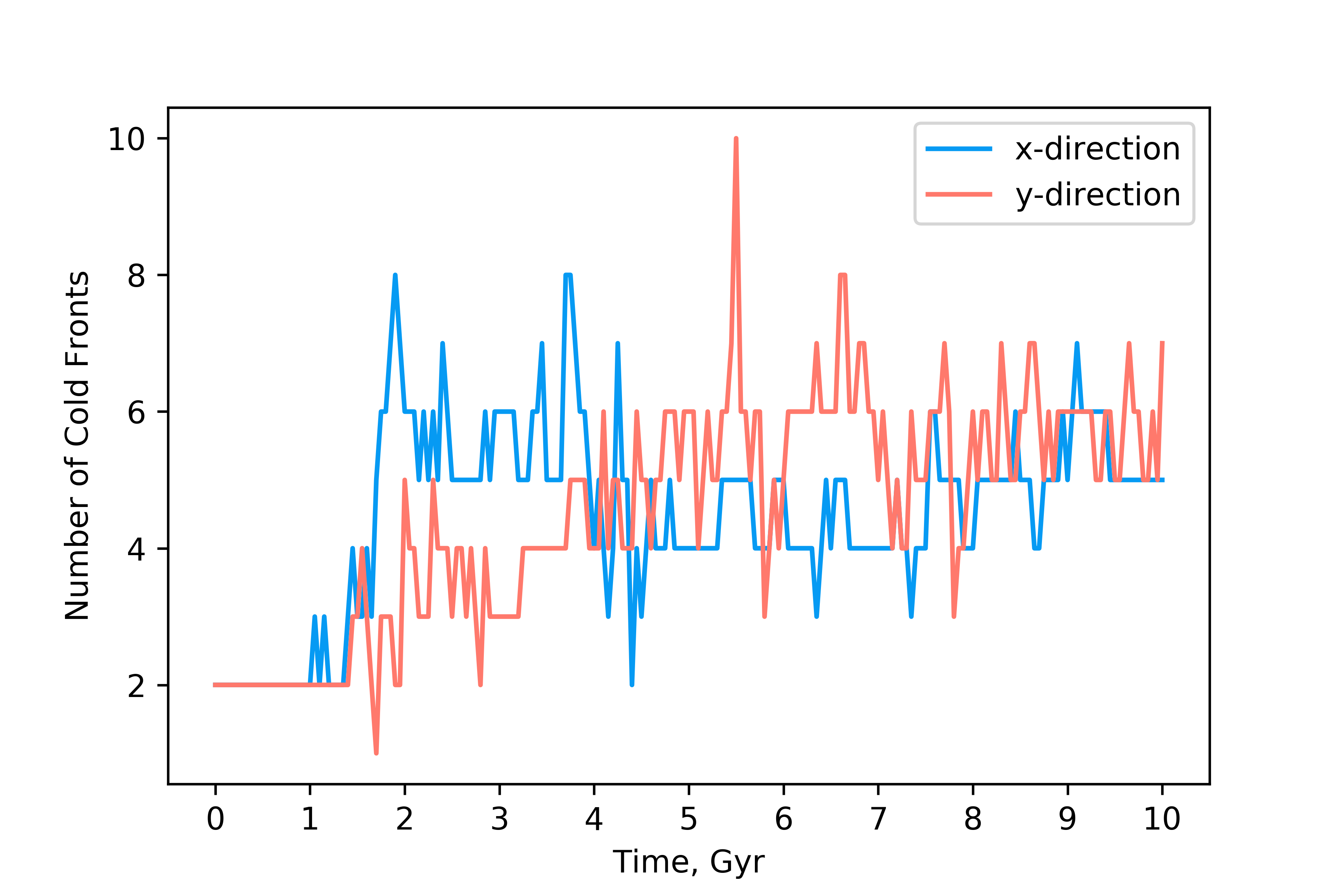}
    \caption{The number of CFs identified in the Binary Merger simulation at 0.5\,Gyr intervals. The blue and red lines represent CFs (with a temperature ratio of at least 1.15) identified in a 30$^{\circ}$ sector centred on the $x$ and $y$ axes, respectively. Criteria for CF identifications are described in Section \ref{section:CF_detection}.}
    \label{fig:binary_results}
\end{figure}

The results of the CF counting algorithm performed on the binary merger simulation are shown in Figure \ref{fig:binary_results}.
First pericentric passage occurs at $\sim$1.5\,Gyr, at which point we see the number of detections increase.
We confirm through visual inspection that this is a combination of both SCFs in the core and MCFs as cluster B is travelling roughly along the $x$-axis and thus through the $x$-axis aligned detection sector.
A natural consequence of this is that detection along the $y$-axis more accurately captures the number of SCFs during this period between the first and second pericentric passages.
Second pericentric passage occurs at $\sim$4.1\,Gyr, at which point we see an inversion in the number of CFs detected along the $x$ and $y$ axes.
This inversion appears to be due to cluster B's trajectory through its approach to second pericentre and beyond being along the $y$-axis.
Beyond second apocentre ($\sim$4.8\,Gyr), we see the number of detected CFs settle to approximately 5-7 along each axis.
The variation during this phase is due to the large scale CFs not being consistently detected (see section \ref{section:CF_detection} for reasons), with little variation seen in the detection at small and intermediate radii (<500\,kpc or 0.3$\,r_{\mathrm{200}}$).
The large CF $\sim$750\,kpc north and west of the centre is detected along the $y$-axis but not the $x$-axis.
There is also a CF $\sim$500\,kpc east of the core which is sporadically detected; we note that the gradient in the temperature profile is consistently detected, but the temperature ratio across the front does not always fulfil the detection criterion.

%%%%%%%%%%%%%% BINARY

%%%%%%%%%%%%%%

 %%%%%%%%%%%%%%%%%%%%%%%%%%%%%%%%%%%%%%%%%%%%% TRIPLES %%%%%%%%%%%%%%%%%%%%%%%%%%%%%%%%%%%%%%%%%%%%%%%%%
 \begin{figure*}
	\includegraphics[width=\textwidth]{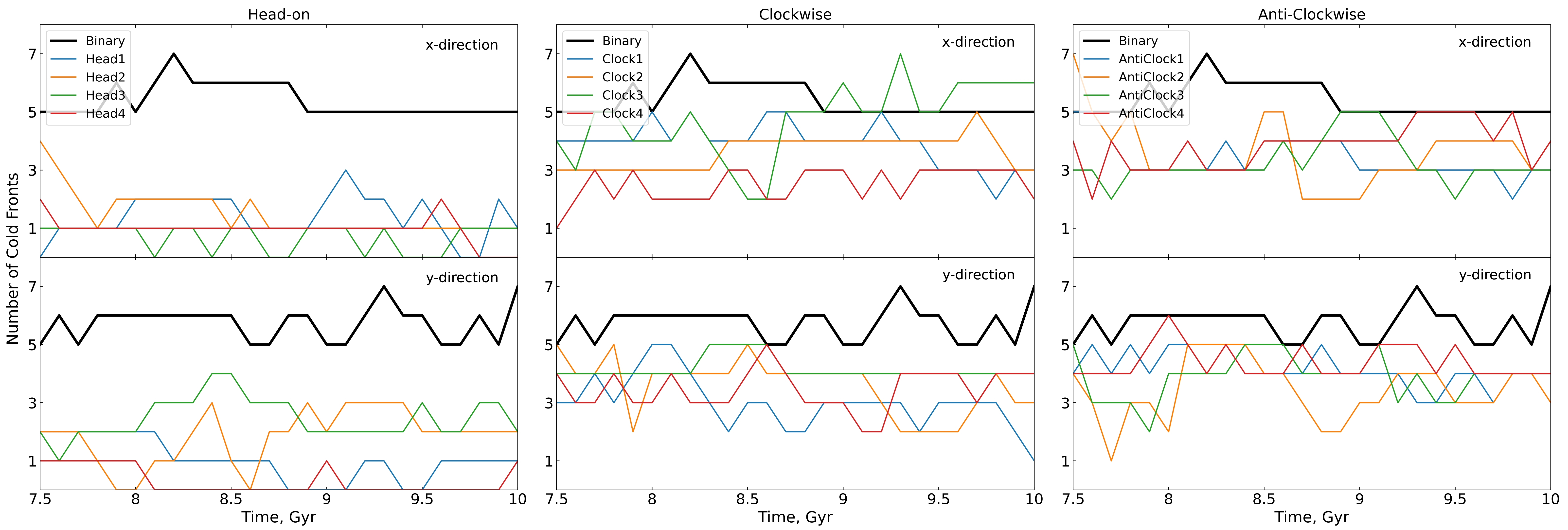}
\caption{The number of detected CFs in all simulations. The black line shows the number of detected CFs in the binary merger simulation, and the coloured lines show results from triple mergers. We see that the number of CFs detected in the binary merger is, in general, greater than in the triple mergers. We see a clear trend that the number of CFs is lower in all triple mergers in which cluster C has collided with the main cluster core head-on. In all off-axis mergers, the number of CFs is only mildly reduced, generally by less than a factor of two.}
\label{fig:CF_numbers}
\end{figure*}

%%%%%%%%%%%%%%%%%%%%%%%%%%%%%%%%%%%%%%%%%%%%% 
\subsubsection{Head-on Mergers}
We have seen from visual inspection (see Section \ref{section:HeadOnVisual}) that the head-on mergers disrupt the cool core of the system, and as such we see no SCFs detected in the inner $\sim$250\,kpc during this phase (7.5-10\,Gyr), with the exception of Head1.
Consistent detections occur, therefore, at or beyond $\sim$200\,kpc (0.12$\,r_{\mathrm{200}}$) towards $t_{\mathrm{max}}$.
Large scale CFs, at $\sim$800\,kpc, are only detected at fortuitous moments when they lie at shallow angles to the detection sector.
Detection of large scale CFs along $y$ is found to be more reliable than detection along $x$, raising the interesting question of whether it is the initial merger that sets the direction of the large scale sloshing, regardless of the second infaller's direction.
We see the number of detected CFs is consistently lower in all head-on simulations when compared to the simple binary merger.
The number of CFs with at least a temperature ratio of 1.15 only drops to zero along \emph{both} $x$ and $y$ at a few timesteps.
We therefore conclude that even a zero-impact parameter merger is not sufficient to destroy all CFs established by a previous merger, but only those at small cluster-centric radii.
Intermediate ($\sim$0.12-0.3 $\,r_{\mathrm{200}}$) and large scale CFs can be detected consistently during the `relaxation phase'. % 7.5-10\,Gyr window. 

 %%%%%%%%%%%%%%%%%%%%%%%%%%%%%%%%%%%%%%%%%%%%% 
\subsubsection{Clockwise and Anti-clockwise Mergers}
We have seen from visual inspection that an off-axis infall of cluster C does not significantly disrupt the cool core of the system.
As such, there is cold gas in the core region that is able to slosh and form small scale SCFs.
The number of detected SCFs in the off-axis mergers ($\sim$3-5) is generally greater than the number detected for the head-on mergers, but is lower than in the binary merger; however, there is no time at which \emph{no} SCFs are detected.
This leads us to conclude that the number SCFs of at least a temperature ratio of 1.15 is resilient to subsequent off-axis minor mergers, though we have seen from visual inspection that the small scale SCFs at $t_{\mathrm{max}}$ are initiated by the infall of cluster C.

%%%%%%%%%%%%%%%%%%%%%%%%%%%%%%%%%%%%%%%%%%%%%%%%%%%%%%%%%%%%%%%%%%%%%%%%%%%%%%%%%%%%%%%%%%%%%%%%%%%%
%%%%%%%%%%%%%%%%%%%%%%%%%%%%%%%%%%%%%%%%%% DISCUSSION %%%%%%%%%%%%%%%%%%%%%%%%%%%%%%%%%%%%%%%%%%%%%%%
%%%%%%%%%%%%%%%%%%%%%%%%%%%%%%%%%%%%%%%%%%%%%%%%%%%%%%%%%%%%%%%%%%%%%%%%%%%%%%%%%%%%%%%%%%%%%%%%%%%%
\section{Discussion}
Our simulations indicate that once sloshing due to a 1:10 minor merger is established, a subsequent minor merger with the same mass ratio is not able to destroy SCFs long-term.
It can only disrupt them short-term for about 1Gyr, between the second merger's first pericentric passage to its first apocentre.
In this section, we discuss potential limits and consequences of the results.

%%%%%%%%%%%%%%%%%%%%%%%%%%%%%%%%%%%%%%%%%%%%%%%%%%%%%%%%%%%%%%%%%%%%%%%%%%%%%%%%%%%%%%%%%%%%%%%%%%%%
\subsection{Simulation setup}

\subsubsection{Parameter space sampling}
\label{section:space_sampling}
\begin{figure}
	\includegraphics[width=\columnwidth]{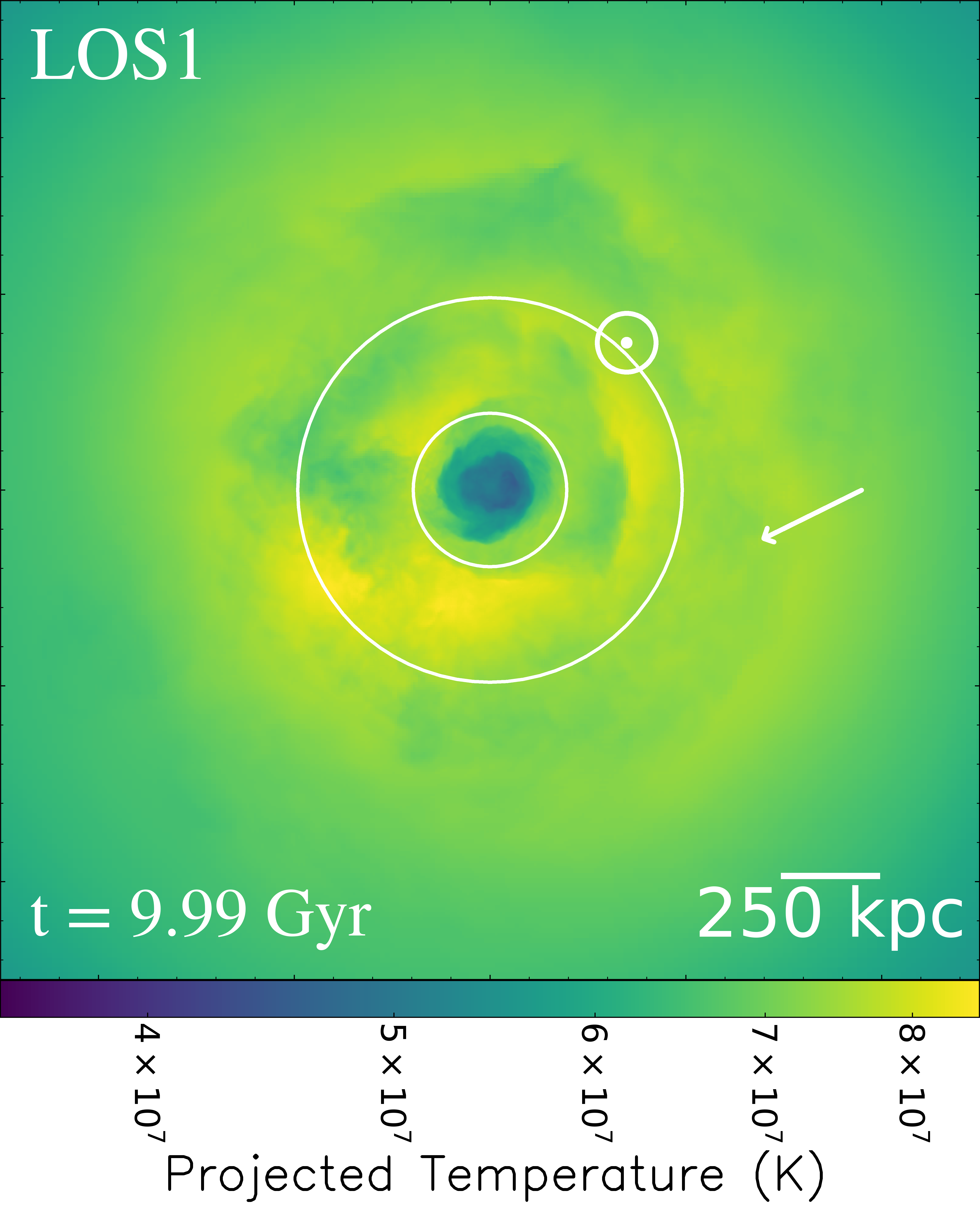}
\caption{Projected temperature weighted by density squared of the LOS1 simulation at $t_{\mathrm{max}}$.The image is 2.5\,Mpc on a side. The arrow represents the trajectory of the first infaller, and the dotted circle represents the trajectory of the second infaller (note that neither annotation is to scale, and should serve only as a guide to the infall directions). Clear SCFs are present in the characteristic spiral pattern, and it is not trivial to distinguish this snapshot from the clockwise or anticlockwise mergers presented in this paper.}
\label{fig:simV}
\end{figure}

The merger parameter space for three clusters is vast.
We have only sampled a small part of this parameter space, namely a constant mass ratio between the three clusters, varying only the impact parameter of the second infalling cluster in the orbital plane of the first merger.
We have performed a supplementary simulation (LOS1) in which cluster C's infall is perpendicular to the initial merger plane and off-axis. 
The state of this simulation at $t_{\mathrm{max}}$ can be seen in Figure \ref{fig:simV}; SCFs are visible in a spiral pattern, and the simulation is qualitatively similar to the clockwise and anti-clockwise simulations.
It is not trivial to distinguish this simulation from the main suite of simulations presented in this paper, lending further weight to the assertion that SCFs should be ubiquitous. 

Even head-on second minor mergers did not erase SCFs established by the first minor merger that had made their way to large radii.
A more massive second infaller may erase these SCFs, but such an event will be rarer than the low-mass case studied here.
The timing of the second merger might also play a role in the distribution of SCFs.
More complete sampling of the parameter space necessitates a separate study due to the size of the parameter space.

%%%%%%%%%%%%%%%%%%%%%%%%%%%%%%%%%%%%%%%%%% WHIM & BOUNDARIES %%%%%%%%%%%%%%%%%%%%%%%%%%%%%%%%%%%%%%%%%%%%%%
\subsubsection{WHIM-like atmosphere and Boundaries}
We wish to keep the simulations as idealised as possible, and as such we must impose certain constraints which have expected consequences.
We have tested the simulations using different combinations of boundary conditions and cluster profiles.
We find that embedding clusters in a WHIM-like atmosphere and in a domain with periodic boundary conditions produces the most realistic results.
As previously stated, the WHIM-like atmosphere acts to pre-truncate the atmospheres of the infalling clusters (B and C), and therefore prevents them from carrying an unrealistic amount of cool gas into the merger.
Clusters are expected to accrete gas from the cosmic web along filaments; cluster A slowly accretes matter from the WHIM-like atmosphere isotropically.
Our simulation suite has also been run with open (`outflow') boundaries in which the gradients of all variables are set to zero at boundaries, allowing matter to flow in or out of the domain.
We find that running the simulation with open boundary conditions for such a long simulation time exacerbates the aforementioned isotropic accretion effect as matter is allowed to flow into the domain, leading to more compression and heating of the ICM.
Even in this scenario, the conclusion of SCF resilience stands.
We therefore implement periodic (`wrap-around') boundaries which effectively act as a universe in miniature and mitigate the accretion of gas onto the cluster.

%%%%%%%%%%%%%%%%%%%%%%%%%%%%%%%%%%%%%%%%%% PHYSICAL EFFECTS %%%%%%%%%%%%%%%%%%%%%%%%%%%%%%%%%%%%%%%%%%%%%%
\subsubsection{Physical effects} 
It is expected that KHIs will develop along CFs and disrupt them \citep{ZuHone2010,Roediger2012a} in the absence of magnetic fields \citep{Lyutikov2006} or viscosity \citep{ZuHone2015} to dampen them.
\citet{ZuHone2010} showed that viscosity suppresses instabilities and turbulence, thus suppressing mixing, which in our case would act to reduce heat transfer to the cool core of cluster A.
This lack of core heating due to mixing is offset in our simulations by the absence of radiative cooling.
However, radiative cooling would help to preserve the cool core and therefore potentially affect the nature of SCFs at small radii.
\citet{ZuHone2011a} found that increased magnetic field strength in the ICM leads to the cool core gas being more resilient to effects of sloshing, and smoother cold fronts.
It is clear that both magnetic fields and viscosity would only make SCFs more resilient to destruction by turbulence and KHI and our conclusions would therefore not be altered by inclusion of these effects.

%%%%%%%%%%%%%%%%%%%%%%%%%%%%%%%%%%%%%%%%%% UBIQUITY %%%%%%%%%%%%%%%%%%%%%%%%%%%%%%%%%%%%%%%%%%%%%%
\subsection{Ubiquity}
Previous work has shown that in relaxed, CC clusters, CFs are likely to be detected \citep{Markevitch2003, GhizzardiRossettiMolendi2010}.
Our simulations of triple mergers with an off-axis second infaller have shown that small and large scale SCFs are present in systems having undergone this type of merger history.
SCFs thus should still be ubiquitous.
Should the merger history of a cluster be such that a head-on collision has disrupted the CC, the simulations have shown that large scale SCFs are still present.
This result sheds some light on systems such as A2142 and A1763 (discussed in sections \ref{section:2142} and \ref{section:1763}, respectively) which lack a strong cool-core but do exhibit cold fronts. 
Of course, since these SCFs are at large radii, they present an observational challenge due to the reduced X-ray emission in cluster outskirts.
However, with the next generation of X-ray telescopes (including the recently launched eRosita), our work suggests we may find that even non-CC clusters host large scale SCFs.
Indeed, a systematic search of archival data may also yield more examples of cold fronts in NCC clusters.
We have also seen that MCFs and SCFs can coexist at many times during a cluster's life.

%%%%%%%%%%%%%%%%%%%%%%%%%%%%%%%%%%%%%%% EXAMPLES %%%%%%%%%%%%%%%%%%%%%%%%%%%%%%%%%%%%%%%%%%%
\subsection{Example Clusters}
Having a single temporal snapshot of a system that is evolving over several Gyr presents a significant challenge to our ability to reconstruct a formation history.
\citet{Roediger2011} deduced the ages of inner SCFs and thus merger history in simulations; \citet{Walker2018} used this same principle to calculate the age of large scale SCFs in the Perseus Cluster.
Our simulations indicate that large-scale SCFs indeed hold clues about longer-term cluster growth.
Detailed analysis on how to distinguish binary and triple mergers and how to determine further merger history will be presented separately (paper in prep.).
In this section, we qualitatively examine some examples from the literature and consider whether their features could indicate a triple merger scenario.
We will consider the projections of temperature weighted by density squared (Figure \ref{fig:TempDens2_Movie_2,5Mpc}) used in the previous analysis to make qualitative comparisons with images found in the literature, noting that confirmation of the scenarios we hypothesise will require further analysis.

%%%%%%%%%%%%%%%%%%%%%%%%%%%%%%%%%%%%%%%
\subsubsection{RXJ2014.8-2430}
RXJ2014.8-2430 ($z=0.138$, $M_{500}=5.38\times 10^{14} M_{\odot}$ \citep{Pratt2010}, $\,r_{\mathrm{200}}=1750$\,Mpc \citep{Croston2008}) is a strong cool core cluster.
\citet{Walker2014} analysed \emph{ROSAT}, \emph{Chandra} and \emph{XMM-Newton} data to reveal two CFs in the central 200\,kpc ($\sim$0.46$\,r_{\mathrm{200}}$) and one to the south at 810\,kpc ($\sim$0.46$\,r_{\mathrm{200}}$).
In our simulations, $\,r_{\mathrm{200}}$ of the primary is 1637\,kpc, and the largest scale CF in the binary merger simulation is $\sim$800\,kpc ($\sim$0.49$\,r_{\mathrm{200}}$) at 10\,Gyr.
The authors suggest that the lack of intermediate fronts indicates that the large-scale CFs arise due to an earlier merger than the inner CFs.
Our simulations suggest that this scenario is entirely plausible.
In particular, several configurations, in which the first and second minor mergers have a non-zero impact parameter produce SCFs which, given appropriate rotation, match this pattern.
For the large scale CF to have had time to reach $\sim$0.46$\,r_{\mathrm{200}}$, we suggest an off-axis minor merger with an age of 7.5-8.5\,Gyr.
%%%%%%%%%%%%%%%%%%%%%%%%%%%%%%%%%%%%%%%

\subsubsection{A2142}
\label{section:2142}
Abell 2142 ($z=0.0909$, $M_{200}=1.25  \times 10^{15} M_{\odot}$, $\,r_{\mathrm{200}}=2.16$\,Mpc \citep{Munari2015}) is an intermediate (between CC and non-CC) cluster which hosts two CFs at small radii, one CF at intermediate radii and one at large radii.
The small/intermediate fronts are at $\sim$300\,kpc ($\sim$0.139$\,r_{\mathrm{200}}$) NW of the centre, $\sim$100\,kpc ($\sim$0.046$\,r_{\mathrm{200}}$) south of centre \citep{Markevitch2000,OwersNulsenCouch2011}, and at 12\,kpc ($\sim$0.0056$\,r_{\mathrm{200}}$) from the gas density peak which connects to the southern front \citep{Johnson2011}.
\citet{Rossetti2013} discovered a fourth CF $\sim$1\,Mpc ($\sim$0.46$\,r_{\mathrm{200}}$) to the southeast of the centre spanning an angular extent of 70$^{\circ}$, corresponding to 1.2\,Mpc.
\citet{Rossetti2013} compared the images of A2142 to the simulations of \citet{Roediger2012a} and \citet{Roediger2012b} and suggested that the fronts in A2142 could have been initiated by an intermediate (i.e. mass ratio of 1:2 or 1:3) merger event.
There is, as in RXJ2014.8-2430, a relative lack of intermediate CFs, suggesting that the large scale CFs are formed by an early off-axis merger, and the small radius CFs are formed by a more recent merger.
\citet{Eckert2017} present a study of the infalling group ($\sim$1.3\,Mpc north of the core), which (by virtue of it exhibiting a slingshot tail in its wake) could be approaching for a second pericentric passage.
The motion of this infalling group is perpendicular to the direction of sloshing, which could imply it caused the small scale sloshing in A2142, but it is unlikely that the CF at 1\,Mpc from the core could have had time to reach this radius if initiated by the infalling group.
This poses an interesting challenge to the assumption that sloshing fronts cannot be generated in isentropic gas (which a NCC approximates).
We see in our simulations with relatively close pericentre distances (e.g. Clock4) that the cool core is weaker than in simulations with a larger pericentre distance. 
It could, therefore, be the case that the infaller responsible for the small scale SCFs in A2142 were caused by a close passing infaller which has significantly disrupted the core but not so much that it is isentropic.
\citet{Einasto2018} showed that A2142 is embedded in a supercluster and suggest that it is the result of several past and ongoing merger events.
Interestingly, the axis of the supercluster (the suggested accretion direction) lies perpendicular to the infalling group of \citet{Eckert2017}.
Given the large scale SCF at $\sim$1\,Mpc, and the ongoing accretion in A2142, this suggests sloshing is indeed resilient to merger activity.
\citet{Venturi2017} found a two-component radio halo in A2142, with the inner component being bounded by the two innermost CFs, and the larger, more diffuse radio emission extending to the largest CF.
It is possible that these two components trace separate merger events in A2142.

%%%%%%%%%%%%%%%%%%%%%%%%%%%%%%%%%%%%%%
\subsubsection{A426 - Perseus}
Perseus ($z=0.017284$ \citep{Hitomi2018}; $M_{200}=6.65 \times 10^{14} M_{\odot}$, $r_{\mathrm{200}}=1.79$\,Mpc \citep{Simionescu2011}) is a relaxed cluster with SCFs in an anti-clockwise spiral pattern originating $\sim$10\,kpc ($\sim$0.0056$\,r_{\mathrm{200}}$) from the centre and terminating with a SCF at $\sim$100\,kpc from the centre \citep{Churazov2003}.
\citet{Churazov2003} state that the chain of galaxies to the west of the BCG (NGC 1275) indicates an ongoing merger with an infaller from the west.
\citet{Simionescu2012} found a surface brightness edge at $\sim$700\,kpc ($\sim$0.39$\,r_{\mathrm{200}}$) east of the cluster centre which they connect to the western edge of the inner SCFs anti-clockwise via a feature north of the centre.
\citet{Walker2020} discovered two large scale fronts (1.2\,Mpc ($\sim$0.67$\,r_{\mathrm{200}}$) and 1.7\,Mpc ($\sim$0.95$\,r_{\mathrm{200}}$)) which they suggest are caused by sloshing initiated $\sim$8.7\,Gyr ago.
The large scale fronts in this case are at considerably larger radii than in our simulations, and so we are hesitant to make a direct comparison with a particular simulation.
We speculate that the old age of the sloshing combined with continuous accretion onto the cluster over many Gyr could move SCFs out to radii as large as observed in Perseus.
However, the apparent lack of CFs at intermediate radii does suggest that this is not a single merger event.
\citet{Zhang2020a} proprosed that the large scale CFs here are in fact shock induced cold fronts (SICFs) caused by a runaway merger shock encountering the accretion shock in the outskirts of the cluster.

%%%%%%%%%%%%%%%%%%%%%%%%%%%%%%%%%%%%%%
\subsubsection{A1763}
\label{section:1763}
Abell 1763 ($z=0.2312$, $M_{200}=1.77\times 10^{15} M_{\odot}$, $\,r_{\mathrm{200}}=1.62$\,Mpc \citep{Rines2013}) is an NCC cluster lying south-west of Abell 1770, with a filament connecting the two clusters.
Abell 1763 exhibits an elongation along the NE-SW axis, consistent with a merger along the filamentary direction.
The system features a double peaked ICM X-ray emission, with the eastern peak coinciding roughly with the BCG \citep{Douglass2018}.
\citet{Haines2018a} identify two $\sim10^{14} M_{\odot}$ subclusters (A1763-g7 and A1763-g9) infalling on Abell 1763 at z=0.237, with A1763-g9 lying $\sim$750\,kpc ($\sim$0.46$\,r_{\mathrm{200}}$) SW of the BCG and A1763-g7 lying $\sim$1.2\,Mpc ($\sim$0.74$\,r_{\mathrm{200}}$) SW of the BCG.
\citet{Douglass2018} observed a spiral pattern in the X-ray surface brightness extending clockwise outward from the western X-ray peak to $\sim$850\,kpc north of the core.
The authors suggest that A1763-g7 is the subcluster responsible for the sloshing spiral, and is returning for an anti-clockwise 2nd pericentric passage.
We see in our simulations that for an $\sim$0.5$\,r_{\mathrm{200}}$ SCF to be present, the sloshing must have been initiated several Gyr ago, by which time the responsible subcluster has merged/ reaches only apocentres of  $\lesssim\,$0.3$\,r_{\mathrm{200}})$.
Therefore we suggest that A1763-g7 is not responsible for the $\sim$850\,kpc SCF, though it is likely responsible for the small-scale SCFs in A1763.

%%%%%%%%%%%%%%%%%%%%%%%%%%%%%%%%%%%%%%%%%%%%%%%%%%%%%%%%%%%%%%%%%%%%%%%%%%%%%%%%%%%%%%%%%%%%%%%%%%%%
%%%%%%%%%%%%%%%%%%%%%%%%%%%%%%%%%%%%%%%%%% CONCLUSIONS %%%%%%%%%%%%%%%%%%%%%%%%%%%%%%%%%%%%%%%%%%%%%%%
%%%%%%%%%%%%%%%%%%%%%%%%%%%%%%%%%%%%%%%%%%%%%%%%%%%%%%%%%%%%%%%%%%%%%%%%%%%%%%%%%%%%%%%%%%%%%%%%%%%%
\section{Conclusions}
We have performed a suite of 13 hydrodynamic + N-body simulations exploring the effects of different triple cluster merger configurations on gas sloshing in the ICM established by a prior off-axis binary minor merger.
We have varied the approach direction and impact parameter of the third cluster to ascertain the resilience of already established sloshing.
We have not yet fully sampled the parameter space, and hence sensitivity to mass ratio and timing of the second merger are not yet explored.

We showed that a second merger leads to a short-term disruption of the SCFs after its first pericentric passage; however during the `relaxation phase' (after its second pericentric passage) we see that the SCFs are present and thus have long-term resilience. We note that it is not trivial to distinguish between the resultant systems at $t_{\mathrm{max}}$.
Our main conclusions are as follows:

1. Cold fronts can always be found in systems that have undergone two minor mergers in sequence, provided the second infalling cluster has a non-zero impact parameter.

2. Sloshing cold front patterns are resilient even to on-axis mergers, with SCFs being temporarily obscured by the second infaller's first pericentric passage and reappearing by the second infaller's second pericentric passage. 

3. Systems containing small ($\sim$0.12$\,r_{\mathrm{200}}$) and large ($\sim$0.5$\,r_{\mathrm{200}}$) scale cold fronts with few or no intermediate fronts are likely to have undergone two off-axis minor mergers in sequence.

4. Clusters with a spiral pattern that can be traced from small to large radii with no gaps are likely to have only undergone a single off-axis binary minor merger, with the pericentre time being calculable from the radius of the largest cold front.

%%%%%%%%%%%%%%%%%%%%%%%%%%%%%%%%%%%%%%%% ACKNOWLEDGEMENTS %%%%%%%%%%%%%%%%%%%%%%%%%%%%%%%%%%%%%%%%%%%%%%
\section*{Acknowledgements}
We thank the anonymous referee for their comments that have improved this paper.
The software used in this work was in part developed by the DOE NNSA-ASC OASCR Flash Center at the University of Chicago.
IV acknowledges the support of University of Hull Astrophysical Data Science Cluster and access to Viper, the University of Hull High Performance Computing Facility, as well as informative conversations with Prof. Brad Gibson.
WF acknowledges support from the Smithsonian Institution, the Chandra High Resolution Camera Project through NASA contract NAS8-03060, and NASA Grants 80NSSC19K0116, GO1-22132X, and GO9-20109X.
CC is funded by the Science and Technology Facilities Council through the University of Hull’s Consolidated Grant ST/R000840/1.
%%%%%%%%%%%%%%%%%%%%%%%%%%%%%%%%%%%%%%%%%%%%%%%%%%%

%%%%%%%%%%%%%%%%%%%%%%%%%%%%%%%%%%%%%%%% DATA AVAILABILITY %%%%%%%%%%%%%%%%%%%%%%%%%%%%%%%%%%%%%%%%%%%%%%
\section*{Data Availability}
The simulation data and code used for analysis from this study are available from IV upon reasonable request.



\bibliographystyle{mnras}
\bibliography{Refs} % if your bibtex file is called example.bib

\begin{thebibliography}{}
\makeatletter
\relax
\def\mn@urlcharsother{\let\do\@makeother \do\$\do\&\do\#\do\^\do\_\do\%\do\~}
\def\mn@doi{\begingroup\mn@urlcharsother \@ifnextchar [ {\mn@doi@}
  {\mn@doi@[]}}
\def\mn@doi@[#1]#2{\def\@tempa{#1}\ifx\@tempa\@empty \href
  {http://dx.doi.org/#2} {doi:#2}\else \href {http://dx.doi.org/#2} {#1}\fi
  \endgroup}
\def\mn@eprint#1#2{\mn@eprint@#1:#2::\@nil}
\def\mn@eprint@arXiv#1{\href {http://arxiv.org/abs/#1} {{\tt arXiv:#1}}}
\def\mn@eprint@dblp#1{\href {http://dblp.uni-trier.de/rec/bibtex/#1.xml}
  {dblp:#1}}
\def\mn@eprint@#1:#2:#3:#4\@nil{\def\@tempa {#1}\def\@tempb {#2}\def\@tempc
  {#3}\ifx \@tempc \@empty \let \@tempc \@tempb \let \@tempb \@tempa \fi \ifx
  \@tempb \@empty \def\@tempb {arXiv}\fi \@ifundefined
  {mn@eprint@\@tempb}{\@tempb:\@tempc}{\expandafter \expandafter \csname
  mn@eprint@\@tempb\endcsname \expandafter{\@tempc}}}

\bibitem[\protect\citeauthoryear{Ascasibar \& Markevitch}{Ascasibar \&
  Markevitch}{2006}]{AscasibarMarkevitch2006}
Ascasibar Y.,  Markevitch M.,  2006, \mn@doi [The Astrophysical Journal]
  {10.1086/506508}, 650, 102

\bibitem[\protect\citeauthoryear{Birnboim, Keshet  \& Hernquist}{Birnboim
  et~al.}{2010}]{Birnboim2010}
Birnboim Y.,  Keshet U.,   Hernquist L.,  2010, \mn@doi [Monthly Notices of the
  Royal Astronomical Society] {10.1111/j.1365-2966.2010.17176.x}, 408, 199

\bibitem[\protect\citeauthoryear{Canning et~al.,}{Canning
  et~al.}{2017}]{Canning2017}
Canning R.~E.,  et~al., 2017, \mn@doi [Monthly Notices of the Royal
  Astronomical Society] {10.1093/mnras/stw2384}, 464, 2896

\bibitem[\protect\citeauthoryear{Cavaliere, Fusco-Fermiano, {A. Cavaliere}  \&
  Fusco-Fermiano}{Cavaliere et~al.}{1976}]{Cavaliere1976}
Cavaliere A.,  Fusco-Fermiano R.,  {A. Cavaliere}  Fusco-Fermiano R.,  1976,
  Astronomy and Astrophysics, 49, 137

\bibitem[\protect\citeauthoryear{Churazov, Forman, Jones  \&
  Bohringer}{Churazov et~al.}{2003}]{Churazov2003}
Churazov E.,  Forman W.,  Jones C.,   Bohringer H.,  2003, \mn@doi [The
  Astrophysical Journal] {10.1086/374923}, 590, 225

\bibitem[\protect\citeauthoryear{Collaboration}{Collaboration}{2018}]{Hitomi2018}
Collaboration H.,  2018, \mn@doi [Publications of the Astronomical Society of
  Japan] {10.1093/pasj/psx138}, 70, 1

\bibitem[\protect\citeauthoryear{Croston et~al.,}{Croston
  et~al.}{2008}]{Croston2008}
Croston J.~H.,  et~al., 2008, \mn@doi [Astronomy and Astrophysics]
  {10.1051/0004-6361:20079154}, 487, 431

\bibitem[\protect\citeauthoryear{Dav{\'{e}} et~al.,}{Dav{\'{e}}
  et~al.}{2001}]{Dave2001}
Dav{\'{e}} R.,  et~al., 2001, \mn@doi [The Astrophysical Journal]
  {10.1086/320548}, 552, 473

\bibitem[\protect\citeauthoryear{Donnert}{Donnert}{2014}]{Donnert2014Gordo}
Donnert J.~M.,  2014, \mn@doi [Monthly Notices of the Royal Astronomical
  Society] {10.1093/mnras/stt2291}, 438, 1971

\bibitem[\protect\citeauthoryear{Douglass, Blanton, Randall, Clarke, Edwards,
  Sabry  \& ZuHone}{Douglass et~al.}{2018}]{Douglass2018}
Douglass E.~M.,  Blanton E.~L.,  Randall S.~W.,  Clarke T.~E.,  Edwards L.
  O.~V.,  Sabry Z.,   ZuHone J.~A.,  2018, \mn@doi [The Astrophysical Journal]
  {10.3847/1538-4357/aae9e7}, 868, 121

\bibitem[\protect\citeauthoryear{Eckert et~al.,}{Eckert
  et~al.}{2017}]{Eckert2017}
Eckert D.,  et~al., 2017, \mn@doi [Astronomy and Astrophysics]
  {10.1051/0004-6361/201730555}, 605, 1

\bibitem[\protect\citeauthoryear{Einasto et~al.,}{Einasto
  et~al.}{2018}]{Einasto2018}
Einasto M.,  et~al., 2018, \mn@doi [Astronomy and Astrophysics]
  {10.1051/0004-6361/201731600}, 610, 1

\bibitem[\protect\citeauthoryear{Fryxell et~al.,}{Fryxell
  et~al.}{2000}]{Fryxell2000}
Fryxell B.,  et~al., 2000, The Astrophysical Journal Supplement Series, 131,
  273

\bibitem[\protect\citeauthoryear{Ghizzardi, Rossetti  \& Molendi}{Ghizzardi
  et~al.}{2010}]{GhizzardiRossettiMolendi2010}
Ghizzardi S.,  Rossetti M.,   Molendi S.,  2010, \mn@doi [Astronomy and
  Astrophysics] {10.1051/0004-6361/200912496}, 516, 1

\bibitem[\protect\citeauthoryear{Haines et~al.,}{Haines
  et~al.}{2018}]{Haines2018a}
Haines C.~P.,  et~al., 2018, \mn@doi [Monthly Notices of the Royal Astronomical
  Society] {10.1093/mnras/sty651}, 477, 4931

\bibitem[\protect\citeauthoryear{Hallman, Skillman, Jeltema, Smith, O'Shea,
  Burns  \& Norman}{Hallman et~al.}{2010}]{Hallman2010}
Hallman E.~J.,  Skillman S.~W.,  Jeltema T.~E.,  Smith B.~D.,  O'Shea B.~W.,
  Burns J.~O.,   Norman M.~L.,  2010, \mn@doi [Astrophysical Journal]
  {10.1088/0004-637X/725/1/1053}, 725, 1053

\bibitem[\protect\citeauthoryear{Hernquist}{Hernquist}{1990}]{Hernquist1990}
Hernquist L.,  1990, \mn@doi [The Astrophysical Journal]
  {10.1017/CBO9781107415324.004}, 356, 359

\bibitem[\protect\citeauthoryear{Johnson}{Johnson}{2011}]{Johnson2011}
Johnson R.~E.,  2011, PhD thesis, Dartmouth College

\bibitem[\protect\citeauthoryear{Kravtsov \& Borgani}{Kravtsov \&
  Borgani}{2012}]{Kravtsov&Borgani2012}
Kravtsov A.~V.,  Borgani S.,  2012, \mn@doi [Annual Review of Astronomy and
  Astrophysics] {10.1146/annurev-astro-081811-125502}, 50, 353

\bibitem[\protect\citeauthoryear{Lyutikov}{Lyutikov}{2006}]{Lyutikov2006}
Lyutikov M.,  2006, \mn@doi [Monthly Notices of the Royal Astronomical Society]
  {10.1111/j.1365-2966.2006.10835.x}, 373, 73

\bibitem[\protect\citeauthoryear{Markevitch \& Vikhlinin}{Markevitch \&
  Vikhlinin}{2007}]{MarkevitchVikhlinin2007}
Markevitch M.,  Vikhlinin A.,  2007, \mn@doi [Physics Reports]
  {10.1016/j.physrep.2007.01.001}, 443, 1

\bibitem[\protect\citeauthoryear{Markevitch et~al.,}{Markevitch
  et~al.}{2000}]{Markevitch2000}
Markevitch M.,  et~al., 2000, \mn@doi [ApJ] {10.1086/309470}, 541, 542

\bibitem[\protect\citeauthoryear{Markevitch, Vikhlinin  \& Forman}{Markevitch
  et~al.}{2003}]{Markevitch2003}
Markevitch M.,  Vikhlinin A.~A.,   Forman W.~R.,  2003, Matter and Energy in
  Clusters of Galaxies ASP Conference Series, 301, 37

\bibitem[\protect\citeauthoryear{Mohr, Mathiesen  \& Evrard}{Mohr
  et~al.}{1999}]{Mohr1999}
Mohr J.~J.,  Mathiesen B.,   Evrard A.~E.,  1999, \mn@doi [The Astrophysical
  Journal] {10.1086/307227}, 517, 627

\bibitem[\protect\citeauthoryear{Munari, Biviano  \& Mamon}{Munari
  et~al.}{2015}]{Munari2015}
Munari E.,  Biviano A.,   Mamon G.~A.,  2015, \mn@doi [Astronomy and
  Astrophysics] {10.1051/0004-6361/201322450e}, 574, 1

\bibitem[\protect\citeauthoryear{Owers, Nulsen, Couch  \& Markevitch}{Owers
  et~al.}{2011}]{OwersNulsenCouch2011}
Owers M.~S.,  Nulsen P.~E.,  Couch W.~J.,   Markevitch M.,  2011, \mn@doi
  [Astrophysical Journal] {10.1088/0004-637X/741/2/122}, 741, 1349

\bibitem[\protect\citeauthoryear{Pratt et~al.,}{Pratt et~al.}{2010}]{Pratt2010}
Pratt G.~W.,  et~al., 2010, \mn@doi [Astronomy and Astrophysics]
  {10.1051/0004-6361/200913309}, 511, 1

\bibitem[\protect\citeauthoryear{Rines, Geller, Diaferio  \& Kurtz}{Rines
  et~al.}{2013}]{Rines2013}
Rines K.,  Geller M.~J.,  Diaferio A.,   Kurtz M.~J.,  2013, \mn@doi
  [Astrophysical Journal] {10.1088/0004-637X/767/1/15}, 767

\bibitem[\protect\citeauthoryear{Roediger \& Zuhone}{Roediger \&
  Zuhone}{2012}]{Roediger2012a}
Roediger E.,  Zuhone J.~A.,  2012, \mn@doi [Monthly Notices of the Royal
  Astronomical Society] {10.1111/j.1365-2966.2011.19794.x}, 419, 1338

\bibitem[\protect\citeauthoryear{Roediger, Br{\"{u}}ggen, Simionescu,
  B{\"{o}}hringer, Churazov  \& Forman}{Roediger et~al.}{2011}]{Roediger2011}
Roediger E.,  Br{\"{u}}ggen M.,  Simionescu A.,  B{\"{o}}hringer H.,  Churazov
  E.,   Forman W.~R.,  2011, \mn@doi [Monthly Notices of the Royal Astronomical
  Society] {10.1111/j.1365-2966.2011.18279.x}, 413, 2057

\bibitem[\protect\citeauthoryear{Roediger, Lovisari, Dupke, Ghizzardi,
  Br{\"{u}}ggen, Kraft  \& Machacek}{Roediger et~al.}{2012}]{Roediger2012b}
Roediger E.,  Lovisari L.,  Dupke R.,  Ghizzardi S.,  Br{\"{u}}ggen M.,  Kraft
  R.~P.,   Machacek M.~E.,  2012, \mn@doi [Monthly Notices of the Royal
  Astronomical Society] {10.1111/j.1365-2966.2011.20287.x}, 420, 3632

\bibitem[\protect\citeauthoryear{Rossetti, Eckert, {De Grandi}, Gastaldello,
  Ghizzardi, Roediger  \& Molendi}{Rossetti et~al.}{2013}]{Rossetti2013}
Rossetti M.,  Eckert D.,  {De Grandi} S.,  Gastaldello F.,  Ghizzardi S.,
  Roediger E.,   Molendi S.,  2013, \mn@doi [Astronomy and Astrophysics]
  {10.1051/0004-6361/201321319}, 556, 1

\bibitem[\protect\citeauthoryear{Sarazin}{Sarazin}{2002}]{Sarazin2002}
Sarazin C.~L.,  2002, \mn@doi [Merging Processes in Galaxy Clusters]
  {10.1007/0-306-48096-4_1}, pp 1--38

\bibitem[\protect\citeauthoryear{Sheardown et~al.,}{Sheardown
  et~al.}{2018}]{Sheardown2018}
Sheardown A.,  et~al., 2018, \mn@doi [The Astrophysical Journal]
  {10.3847/1538-4357/aadc0f}, 865, 118

\bibitem[\protect\citeauthoryear{Sheardown et~al.,}{Sheardown
  et~al.}{2019}]{Sheardown2019}
Sheardown A.,  et~al., 2019, \mn@doi [The Astrophysical Journal]
  {10.3847/1538-4357/ab0c06}, 874, 112

\bibitem[\protect\citeauthoryear{Simionescu et~al.,}{Simionescu
  et~al.}{2011}]{Simionescu2011}
Simionescu A.,  et~al., 2011, Science, 331, 1576

\bibitem[\protect\citeauthoryear{Simionescu et~al.,}{Simionescu
  et~al.}{2012}]{Simionescu2012}
Simionescu A.,  et~al., 2012, \mn@doi [Astrophysical Journal]
  {10.1088/0004-637X/757/2/182}, 757

\bibitem[\protect\citeauthoryear{Springel, {Di Matteo}  \& Hernquist}{Springel
  et~al.}{2005}]{Springel&Hernquist&DiMatteo2005a}
Springel V.,  {Di Matteo} T.,   Hernquist L.,  2005, \mn@doi [Monthly Notices
  of the Royal Astronomical Society] {10.1111/j.1365-2966.2005.09238.x}, 361,
  776

\bibitem[\protect\citeauthoryear{Tittley \& Henriksen}{Tittley \&
  Henriksen}{2005}]{TittleyHenriksen2005}
Tittley E.~R.,  Henriksen M.,  2005, \mn@doi [The Astrophysical Journal]
  {10.1086/425952}, 618, 227

\bibitem[\protect\citeauthoryear{Turk, Smith, Oishi, Skory, Skillman, Abel  \&
  Norman}{Turk et~al.}{2011}]{Turk2011}
Turk M.~J.,  Smith B.~D.,  Oishi J.~S.,  Skory S.,  Skillman S.~W.,  Abel T.,
  Norman M.~L.,  2011, \mn@doi [Astrophysical Journal, Supplement Series]
  {10.1088/0067-0049/192/1/9}, 192

\bibitem[\protect\citeauthoryear{Venturi et~al.,}{Venturi
  et~al.}{2017}]{Venturi2017}
Venturi T.,  et~al., 2017, \mn@doi [Astronomy and Astrophysics]
  {10.1051/0004-6361/201630014}, 603, 1

\bibitem[\protect\citeauthoryear{Vikhlinin, Markevitch  \& Murray}{Vikhlinin
  et~al.}{2001}]{Vikhlinin2001}
Vikhlinin A.,  Markevitch M.,   Murray S.~S.,  2001, \mn@doi [The Astrophysical
  Journal] {10.1086/320078}, 551, 160

\bibitem[\protect\citeauthoryear{Vitvitska, Klypin, Kravtsov, Wechsler, Primack
   \& Bullock}{Vitvitska et~al.}{2002}]{Vitvitska2002}
Vitvitska M.,  Klypin A.~A.,  Kravtsov A.~V.,  Wechsler R.~H.,  Primack J.~R.,
   Bullock J.~S.,  2002, The Astrophysical Journal, 581, 799

\bibitem[\protect\citeauthoryear{Walker, Fabian  \& Sanders}{Walker
  et~al.}{2014}]{Walker2014}
Walker S.~A.,  Fabian A.~C.,   Sanders J.~S.,  2014, \mn@doi [Monthly Notices
  of the Royal Astronomical Society: Letters] {10.1093/mnrasl/slu040}, 441, 31

\bibitem[\protect\citeauthoryear{Walker, Zuhone, Fabian  \& Sanders}{Walker
  et~al.}{2018}]{Walker2018}
Walker S.~A.,  Zuhone J.,  Fabian A.,   Sanders J.,  2018, \mn@doi [Nature
  Astronomy] {10.1038/s41550-018-0401-8}, 2, 292

\bibitem[\protect\citeauthoryear{Walker, Mirakhor, ZuHone, Sanders, Fabian  \&
  Diwanji}{Walker et~al.}{2020}]{Walker2020}
Walker S.~A.,  Mirakhor M.~S.,  ZuHone J.,  Sanders J.~S.,  Fabian A.~C.,
  Diwanji P.,  2020

\bibitem[\protect\citeauthoryear{Zhang, Churazov, Dolag, Forman  \&
  Zhuravleva}{Zhang et~al.}{2020}]{Zhang2020a}
Zhang C.,  Churazov E.,  Dolag K.,  Forman W.~R.,   Zhuravleva I.,  2020,
  \mn@doi [Monthly Notices of the Royal Astronomical Society: Letters]
  {10.1093/mnrasl/slaa147}, 498, L130

\bibitem[\protect\citeauthoryear{Zinger, Dekel, Birnboim, Nagai, Lau  \&
  Kravtsov}{Zinger et~al.}{2018}]{Zinger2018}
Zinger E.,  Dekel A.,  Birnboim Y.,  Nagai D.,  Lau E.,   Kravtsov A.~V.,
  2018, \mn@doi [Monthly Notices of the Royal Astronomical Society]
  {10.1093/mnras/sty136}, 476, 56

\bibitem[\protect\citeauthoryear{ZuHone}{ZuHone}{2011}]{ZuHone2011}
ZuHone J.~A.,  2011, \mn@doi [Astrophysical Journal]
  {10.1088/0004-637X/728/1/54}, 728

\bibitem[\protect\citeauthoryear{ZuHone \& Roediger}{ZuHone \&
  Roediger}{2016}]{ZuHone2016a}
ZuHone J.~A.,  Roediger E.,  2016, \mn@doi [Journal of Plasma Physics]
  {10.1017/S0022377816000544}, 82

\bibitem[\protect\citeauthoryear{ZuHone, Markevitch  \& Johnson}{ZuHone
  et~al.}{2010}]{ZuHone2010}
ZuHone J.~A.,  Markevitch M.,   Johnson R.~E.,  2010, \mn@doi [Astrophysical
  Journal] {10.1088/0004-637X/717/2/908}, 717, 908

\bibitem[\protect\citeauthoryear{ZuHone, Markevitch  \& Lee}{ZuHone
  et~al.}{2011}]{ZuHone2011a}
ZuHone J.~A.,  Markevitch M.,   Lee D.,  2011, \mn@doi [Astrophysical Journal]
  {10.1088/0004-637X/743/1/16}, 743

\bibitem[\protect\citeauthoryear{ZuHone, Kunz, Markevitch, Stone  \&
  Biffi}{ZuHone et~al.}{2015}]{ZuHone2015}
ZuHone J.~A.,  Kunz M.~W.,  Markevitch M.,  Stone J.~M.,   Biffi V.,  2015,
  \mn@doi [Astrophysical Journal] {10.1088/0004-637X/798/2/90}, 798

\makeatother
\end{thebibliography}

%%%%%%%%%%%%%%%%%%%%%%%%%%%%%%%%%%%%%%%%%%%%%%%%%%

%%%%%%%%%%%%%%%%% APPENDICES %%%%%%%%%%%%%%%%%%%%%
\appendix
\section{Appendix A} \label{section:Appendix A}

%%%%%%%%%%%%%%%%%%%%%%%%%%%%%%%% TRIPLES
\begin{landscape}
\begin{figure}
   \includegraphics[width=\textheight]{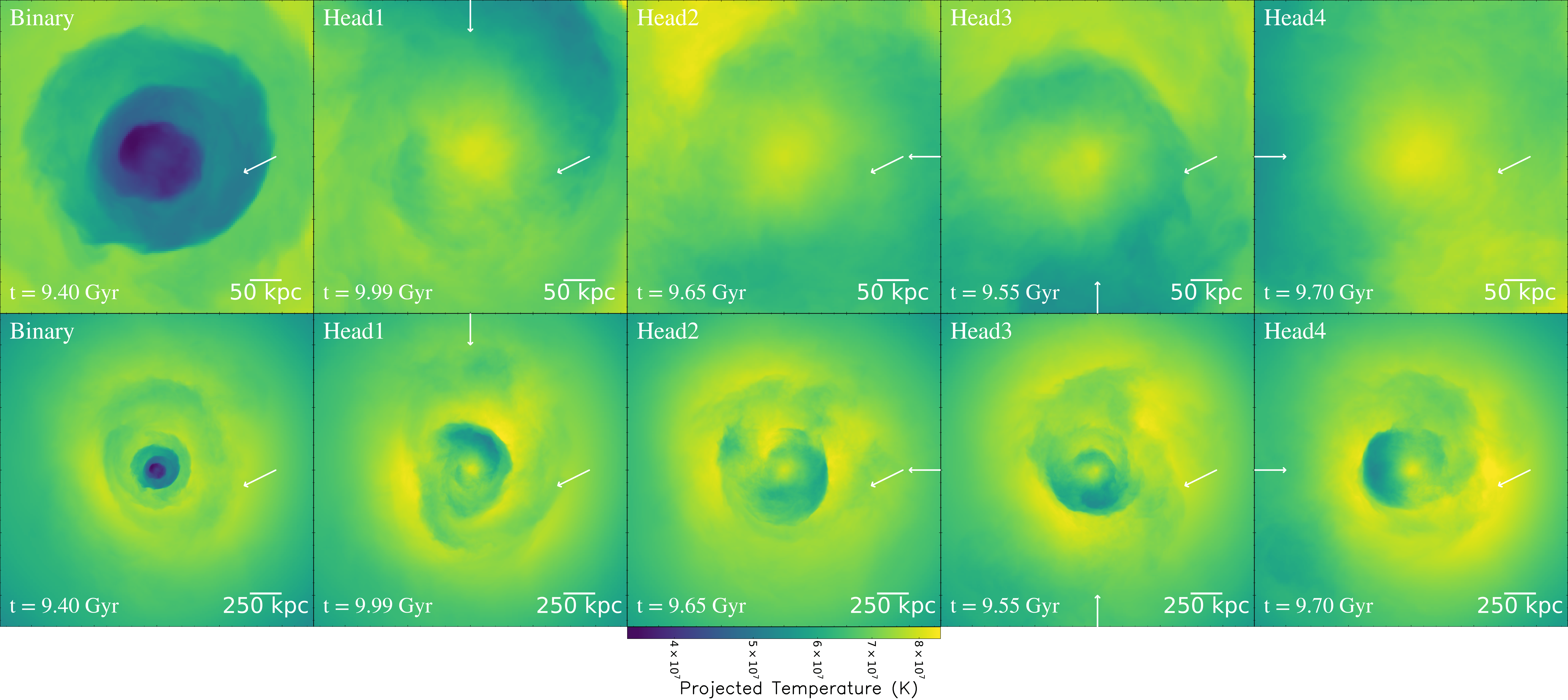}  %this is a placeholder image for A1_Head_2ndInfaller.mov
    \caption{Movie (the movie can be viewed at https://youtu.be/q-CqnaGtg0E)of projections of temperature with density squared weighting through the $x-y$ plane for the head-on simulations with the binary merger on the left column for reference. All panels are centred on the minimum potential of the domain (the centre of cluster A). The top row panels are 500\,kpc on a side and the bottom row panels are 2.5\,Mpc on a side. The individual simulations have been staggered such that the first pericentre of cluster C occurs at the same time in the movie. The arrows have the same meaning as in previous figures. We clearly see that the head-on merger has disrupted the cool-core of cluster A but that cold fronts are still visible outside the core region.}
    \label{fig:HeadOn_Movie}
\end{figure}
\end{landscape}

\begin{landscape}
\begin{figure}
   \includegraphics[width=\textheight]{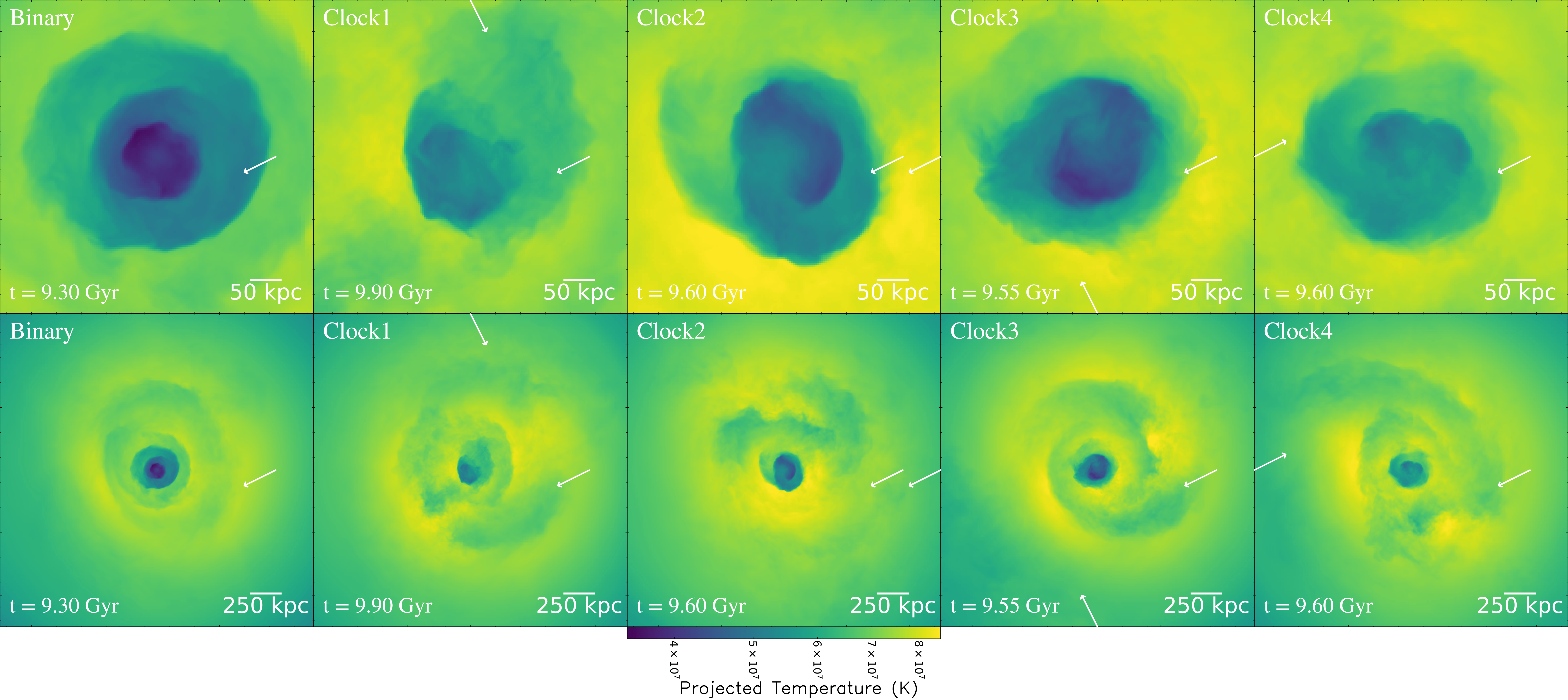}  %this is a placeholder image for A2_Clock_2ndInfaller.mov
    \caption{Movie (the movie can be viewed at https://youtu.be/ptxtqtfvlSU) of projections of temperature with density squared weighting through the $x-y$ plane for the clockwise simulations with the binary merger on the left column for reference. All panels are centred on the minimum potential of the domain (the centre of cluster A). The top row panels are 500\,kpc on a side and the bottom row panels are 2.5\,Mpc on a side. The individual simulations have been staggered such that the first pericentre of cluster C occurs at the same time in the movie. The arrows have the same meaning as in previous figures. We see that the clockwise mergers do not disrupt the cold fronts that have already travelled out of the core region of cluster A and that new sloshing is initiated by cluster C.}
    \label{fig:Clockwise_Movie}
\end{figure}
\end{landscape}

\begin{landscape}
\begin{figure}
   \includegraphics[width=\textheight]{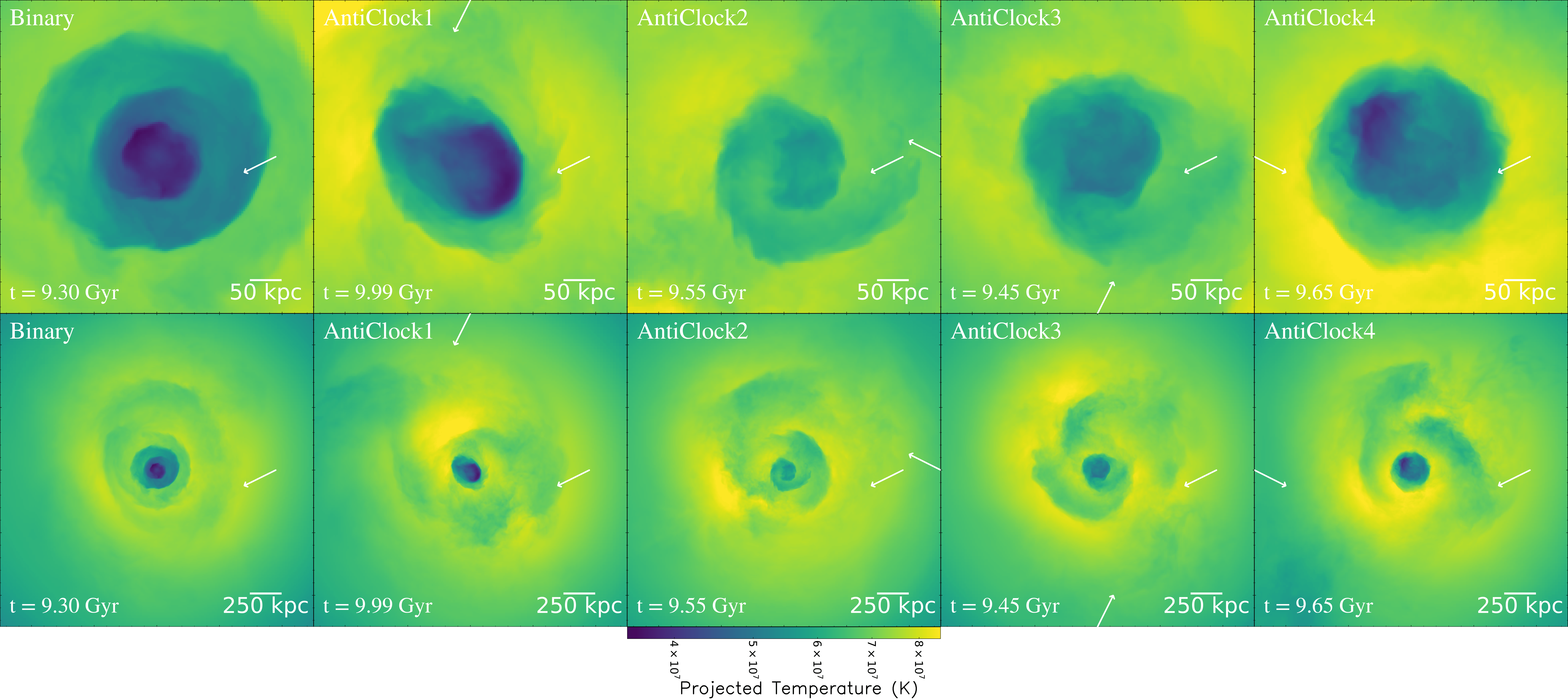} %this is a placeholder image for A3_AntiClock_2ndInfaller.mov
    \caption{Movie (the movie can be viewed at https://youtu.be/fNv38D89Phk) of projections of temperature with density squared weighting through the $x-y$ plane for the anti-clockwise simulations with the binary merger on the left column for reference. All panels are centred on the minimum potential of the domain (the centre of cluster A). The top row panels are 500\,kpc on a side and the bottom row panels are 2.5\,Mpc on a side. The individual simulations have been staggered such that the first pericentre of cluster C occurs at the same time in the movie. The arrows have the same meaning as in previous figures. We see that the anti-clockwise mergers do not disrupt the cold fronts that have already travelled out of the core region of cluster A and that new sloshing is initiated by cluster C.}
    \label{fig:AntiClock_Movie}
\end{figure}
\end{landscape}
%%%%%%%%%%%%%%%%%%%%%%%%%%%%%%%%%%%%%

\begin{landscape}
\begin{figure}
   \includegraphics[width=\textheight]{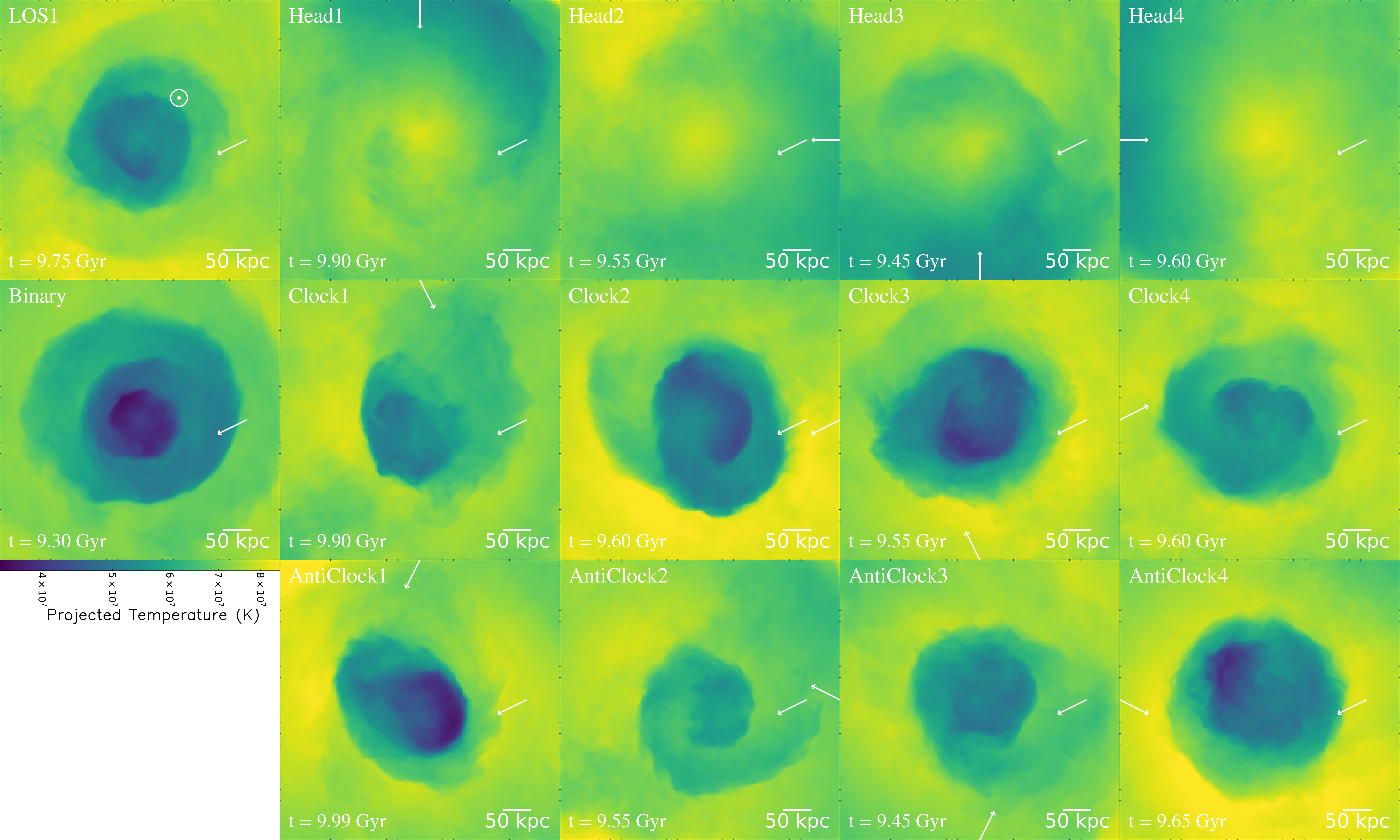} %this is a placeholder image for A4_0,5Mpc_2ndInfallLater.mov
    \caption{Movie (the movie can be viewed at https://youtu.be/0bICdEy1F3M) of projections of temperature with density squared weighting through the $x-y$ plane for all simulations. Each panel is 0.5\,Mpc on a side and the simulation times have been altered such that the simulations simultaneously show cluster C's first pericentric passage. The rows are comprised, respectively, of the head-on, clockwise and anti-clockwise simulations; the columns represent approach direction of cluster C. The leftmost column is the binary merger and supplementary simulation (discussed in Sec \ref{section:space_sampling}) with the colour scale for all simulations. The arrows and dotted circle have the same meaning as in previous figures.}
    \label{fig:TempDens2_Movie_0,5Mpc}
\end{figure}
\end{landscape}

\begin{landscape}
\begin{figure}
   \includegraphics[width=\textheight]{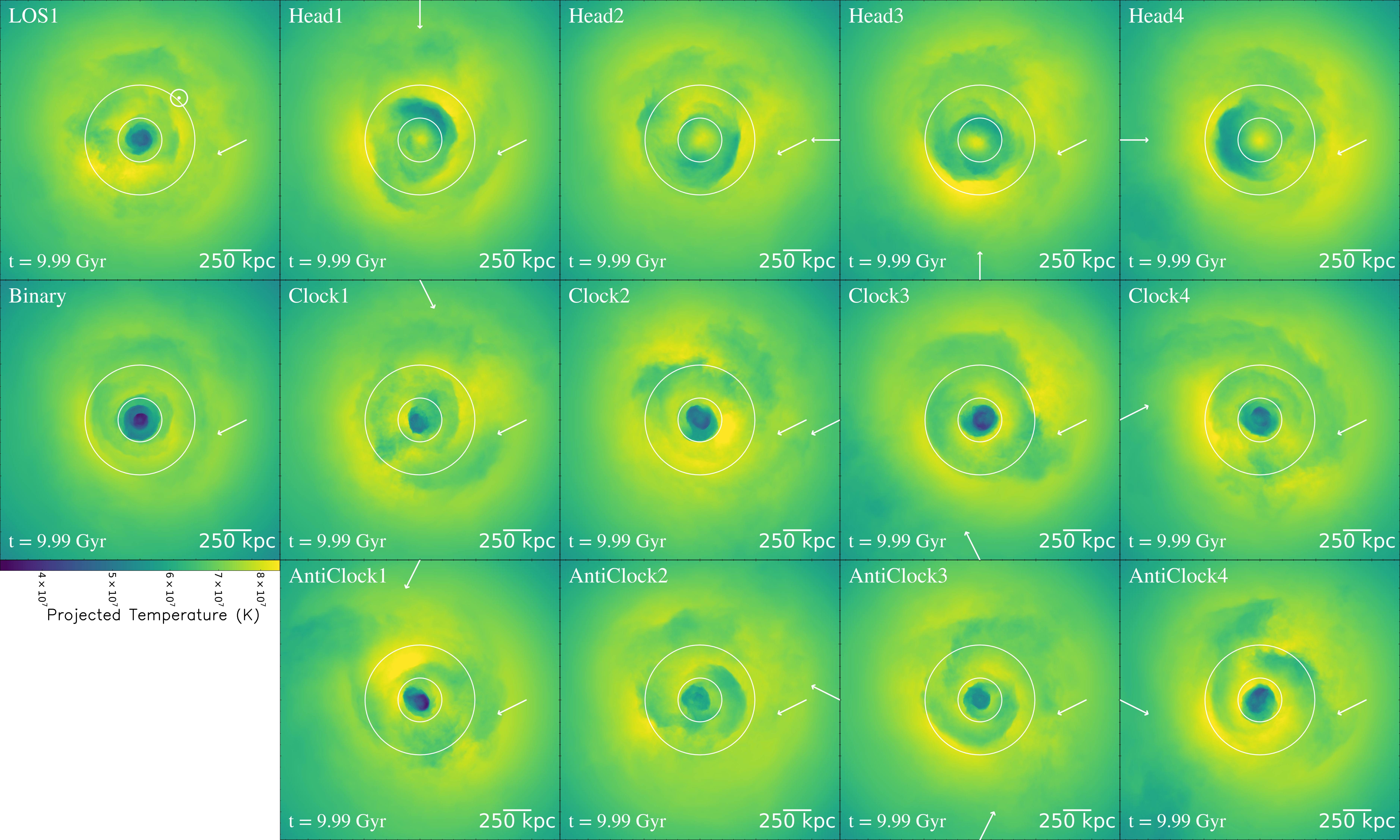} 
       \caption{Projection of temperature with density squared weighting through the $x-y$ plane for all simulations at the maximum simulation time for each simulation. Each panel is 2.5\,Mpc on a side and the annotations have the same meaning as in previous figures.}
    \label{fig:TempDens2_Movie_0,5Mpc}
\end{figure}
\end{landscape}

%%%%%%%%%%%%%%%%%%%%%%%%%%%%%%%%%%%%%%%%%%%%%%%%%%
% Don't change these lines
%\bsp	% typesetting comment
%\label{lastpage}
\end{document}